%% file: ms.tex
\documentclass[dvipdfmx]{pasj01}
\draft 
\Received{$\langle$2016-09-09$\rangle$}
\Accepted{$\langle$accepted date$\rangle$}
\Published{$\langle$publication date$\rangle$}

\usepackage{color}

\usepackage{ulem}

\newcommand{\feoh}{\textrm{[Fe/H]}}
\newcommand{\cfe}{\textrm{[C/Fe]}}
\newcommand{\mgfe}{\textrm{[Mg/Fe]}}
\newcommand{\mgh}{\textrm{[Mg/H]}}
\newcommand{\xbyh}{\textrm{[X/H]}}
\newcommand{\xbyfe}{\textrm{[X/Fe]}}

\newcommand{\afe}{\textrm{[$\alpha$/Fe]}}
\newcommand{\logeps}{\ensuremath{\log \epsilon (\textrm{X})}}

\newcommand{\msun}{\ensuremath{M_{\odot}}}

\newcommand{\teff}{\ensuremath{T_{\textrm {eff}}}}
\newcommand{\logg}{\ensuremath{\log {\textrm g}}}
\newcommand{\hyp}{\ensuremath{\, \mathchar`- \,}}

\newcommand{\rit}{{\it r}}
\newcommand{\sit}{{\it s}}
\newcommand{\cemps}{CEMP-{\it s}}
\newcommand{\cempno}{CEMP-no}

\newcommand{\mejeu}{M_{\rm ej, Eu}}
\newcommand{\mejba}{M_{\rm ej, Ba}}
\newcommand{\mejfe}{M_{\rm ej, Fe}}
\newcommand{\mismeu}{M_{\rm ISM, Eu}}
\newcommand{\mismba}{M_{\rm ISM, Ba}}
\newcommand{\mismfe}{M_{\rm ISM, Fe}}
\newcommand{\xeusun}{X_{\rm Eu, \odot}}
\newcommand{\xbasun}{X_{\rm Ba, \odot}}
\newcommand{\xfesun}{X_{\rm Fe, \odot}}
\newcommand{\euba}{[{\rm Eu}/{\rm Ba}]}
\newcommand{\eufe}{[{\rm Eu}/{\rm Fe}]}
\newcommand{\bafe}{[{\rm Ba}/{\rm Fe}]}
\newcommand{\bah}{[{\rm Ba}/{\rm H}]}

\begin{document}

\title{Stellar Abundances for Galactic Archaeology Database IV - Compilation of Stars in Dwarf Galaxies}
\author{Takuma Suda$^{1}$}%
\author{Jun Hidaka$^{2}$}%
\author{Wako Aoki$^{3}$}%
\author{Yutaka Katsuta$^{4}$}%
\author{Shimako Yamada$^{4}$}%
\author{Masayuki Y. Fujimoto$^{5}$}%
\author{Yukari Ohtani$^{6}$}%
\author{Miyu Masuyama$^{1,7}$}%
\author{Kazuhiro Noda$^{1,7}$}%
\author{Kentaro Wada$^{1,7}$}%
\altaffiltext{1}{Research Center for the Early Universe, The University of Tokyo,
 7-3-1 Hongo, Bunkyo-ku, Tokyo 113-0033, Japan }
\altaffiltext{2}{School of Science and Engineering, Meisei University, 2-1-1 Hodokubo, Hino, Tokyo 191-0042, Japan}
\altaffiltext{3}{National Observatory of Japan,
 Osawa 2-21-1, Mitaka, Tokyo 181-8588, Japan }
\altaffiltext{4}{Department of Cosmosciences, Hokkaido University,
 Kita 10 Nishi 8, Kita-ku, Sapporo 060-0810, Japan }
\altaffiltext{5}{Faculty of Engineering, Hokkai-Gakuen University,
 Minami 26 Nishi 11, Chuo-ku, Sapporo 064-0926, Japan }
\altaffiltext{6}{Center for Computational Astrophysics, National Observatory of Japan,
 Osawa 2-21-1, Mitaka, Tokyo 181-8588, Japan }
\altaffiltext{7}{Department of Astronomy, Graduate School of Science, The University of Tokyo,
 7-3-1 Hongo, Bunkyo-ku, Tokyo 113-0033, Japan }
\email{suda@resceu.s.u-tokyho.ac.jp}


\maketitle

\begin{abstract}
We have constructed the database of stars in local group galaxies using the extended version of the SAGA (Stellar Abundances for Galactic Archaeology) database that contains stars in 24 dwarf spheroidal galaxies and ultra faint dwarfs.
The new version of the database includes more than $4500$ stars in the Milky Way, by removing the previous metallicity criterion of $\feoh \leq -2.5$, and more than $6000$ stars in the local group galaxies.
We examined a validity of using a combined data set for elemental abundances.
We also checked a consistency between the derived distances to individual stars and those to galaxies in the literature values.
Using the updated database, the characteristics of stars in dwarf galaxies are discussed.
Our statistical analyses of $\alpha$-element abundances show that the change of the slope of the \afe\ relative to \feoh\ (so-called ``knee'') occurs at $\feoh = -1.0 \pm 0.1$ for the Milky Way.
The knee positions for selected galaxies are derived by applying the same method.
Star formation history of individual galaxies are explored using the slope of the cumulative metallicity distribution function.
Radial gradients along the four directions are inspected in six galaxies where we find no direction dependence of metallicity gradients along the major and minor axes.
The compilation of all the available data shows a lack of \cemps\ population in dwarf galaxies, while there may be some \cempno\ stars at $\feoh \lesssim -3$ even in the very small sample.
The inspection of the relationship between Eu and Ba abundances confirms an anomalously Ba-rich population in Fornax, which indicates a pre-enrichment of interstellar gas with \rit-process elements.
We do not find any evidence of anti-correlations in O-Na and Mg-Al abundances, which characterises the abundance trends in the Galactic globular clusters.
\end{abstract}

\section{Introduction}

Satellite galaxies in the local group are thought to be survivors of the building blocks of the Milky Way (MW) and may contain important information for its formation.
They provide opportunities for inspecting similarity and difference among satellite galaxies and our Galaxy, in terms of chemodynamical evolution (see e.g., \cite{Freeman2002}).
To understand the environment of star formation and galaxy formation, stellar elemental abundances are useful tools, providing us with much information on the chemical history of MW and the local group through the comparisons of models with observations.
In particular, extremely metal-poor stars in the Galactic halo help to understand the early phase of chemical evolution, stellar evolution, and nucleosynthesis (see e.g., \cite{Frebel2015}).
The success of the so called near field cosmology or chemical tagging has been extended to stars in the local group galaxies in the last decade with the improvement of facilities for spectroscopic studies.
Among them, dwarf galaxies are small but plentiful in the local group, and close enough to obtain elemental abundances of individual stars.
The most abundant and well-studied ones are dwarf spheroidal galaxies (dSphs) and ultra-faint dwarf galaxies (UFDs).
They do not show evidence of active star formation at present but are thought to have a complex star formation history from galaxy to galaxy \citep{Grebel2003}.

Due to the limiting magnitudes to obtain spectroscopic data, stellar elemental abundances of dwarf galaxies are available only for red giants.
There are two ways of data production in the current observational campaigns for abundance analyses.
One is a massive data production with medium-resolution spectroscopy using a multi-fiber spectrograph such as the Keck DEIMOS and the VLT GIRAFFE in the Medusa mode.
The other is an accumulation of a few data with high-resolution spectroscopy using a high dispersion spectrograph, as in the case of the Galactic halo stars.
In the former case, the abundances are determined by empirical formulae for the limited number of elements, not by spectral analyses based on stellar atmospheric parameters.
In the latter case, the procedure is the same as in the Galactic stars, and hence, we can safely compare the abundance patterns between the target galaxy and MW.
The largest project of collecting data of stellar abundances in dSphs is the DART project \citep{Tolstoy2006}.
It produces many papers reporting elemental abundances using the medium- and high-resolution spectrographs on VLT.
On the other hand, the most abundant data have been produced by Kirby and his collaborators using the Keck DEIMOS.
In particular, \citet{Kirby2010} reported abundances of iron and $\alpha$-element in about 3,000 stars.
For the case of iron abundance only, \citet{Koch2006} published data for more than one thousand stars.

Thanks to these efforts, accumulated data of stars in the dwarf galaxies enable us to compare with those in MW.
For example, many studies focus on the comparison of \afe\ with respect to \feoh, trying to understand the contribution of AGB stars, Type Ia supernovae, and Type II supernovae to the chemical enrichment \citep{Shetrone2003,Monaco2005,Battaglia2006,Sbordone2007,Cohen2009,Aoki2009,Feltzing2009,Norris2010b,Gilmore2013,Hendricks2014a,Lemasle2014,Kirby2015,Fabrizio2015}.
The studies with high-resolution spectra help to understand the chemical enrichment history of iron-group elements \citep{Sadakane2004,Koch2008a,Tafelmeyer2010,Simon2010,North2012} and neutron-capture elements \citep{Letarte2010,Frebel2010,Honda2011,Venn2012,Ishigaki2014,Jablonka2015,Skuladottir2015,Tsujimoto2015,Ural2015,Ji2016b,Roederer2016}.
The detailed abundances are derived with medium-resolution spectra \citep{Bosler2007,Starkenburg2013}, which can be also comparable with high-resolution ones.
On the other hand, it is not easy to compare data from various papers due to the difficulty in compiling data from individual papers, which has been increasing in the last decade.
It is also difficult to check the identity of objects in different papers since most of the papers use different names for the same object.
In comparing the abundance data, normalisation with adopted solar abundance can be a concern if different papers employ different standard solar abundances.

The SAGA (Stellar Abundances for Galactic Archaeology) database \footnote{http://sagadatabase.jp} \citep{Suda2008,Suda2011,Yamada2013} helps to solve these problems since it processes the data automatically once the papers and the data therein are registered in the database.
The database contains the data of metal-poor stars in MW, especially covering almost all the known extremely metal-poor (EMP) stars in the Galactic halo.
In this paper, we report an extension of the database that includes stars in the local group together with some metal-rich stars in MW as a comparable counterpart.
We also report some improvements in the treatment of data and in the web interface to retrieve and plot the data.

This paper is organised as follows.
The next section is devoted to the overview of the database, which includes how the database works, what are the criteria for data compilation, and what has been updated since the previous work.
Some notes and comments on individual galaxies are described in \S~3
In \S~4, users of the database are cautioned about the accuracy of derived surface gravities in the original data, although its impact on final abundances will not be significant.
Similarity and difference among dwarf galaxies are discussed in \S~5, followed by summary in \S~6.

\section{Overview of the Database}

The SAGA database for the local group galaxies consists of available data for stars in dwarf galaxies in the local group.
The data are taken from literature and are compiled for observational and theoretical studies.
We have collected papers that report at least one elemental abundance in the samples.
We exclude stars in globular clusters that are often listed together with dSph stars as a comparison sample in the current version of the database.

The quantities compiled in the database are almost the same as those for EMP stars in MW.
The data include bibliographic data, a log of observations, the positions of the stars, stellar parameters, photometric data, equivalent widths, binary parameters, adopted solar abundances, and elemental abundances.
In the current version of the database, binary parameters in any dSphs are not included due to the lack of information in the literature.
To identify the galaxy to which stars belong, we added a record for the names of member galaxies.
All the stars, including MW stars, have assigned galaxy names in the new version of the database.
So far, there are 25 dSph and ultra-faint dwarf galaxies registered in the library of the database.

Currently there are two public versions of the database.
One is the data for MW stars only, which is the same as the original version.
It is still useful because the sample can be classified by abundances and evolutionary status.
The other is the combined database that contains all the stars in MW and dwarf galaxies.
In the combined version of the database, we can compare the characteristics of stars in any choice of dwarf galaxies with those in MW.
In this dataset, we do not divide the sample by metallicity, carbon abundance, or evolutionary status as in the case of the stars in MW, for which stars are divided by $\feoh = -2.5$, $\cfe = 0.7$, or by stellar parameters to distinguish whether the star is a red giant or a dwarf.

The current version of the database includes 6039 unique stars in dwarf galaxies with additional 2524 stars that have independent measurements of abundances (see the next section.) from 78 papers.
We have covered all the papers, which date back to 1998, regarding the determination of stellar abundances in the local group.
The total number of abundance data is 24765 for 38 elements.

The database consists of four sub-systems: the reference management sub-system, the data compilation sub-system, the data registration sub-system, and the data retrieval sub-system.
The details of these subsystems are described in \citet[(hereafter Paper I)]{Suda2008}.
Here we briefly describe the main features and the detail of the updates since the previous version.

The reference management subsystem is a repository for the relevant papers.
The list of papers includes not only papers reporting abundances, but also papers dealing with the confirmation of membership and analyses of AGB stars, the latter of which have not yet compiled.
As of 2017 May, a total of 144 papers is registered for the category of the local group ({\it cf.}, 340 papers for MW), 78 of which are compiled.
As noted above, the only requirement for compilation is that stars in the local group galaxies are neither AGB stars nor stars in globular clusters.
We do not need any modifications to this subsystem and are able to simply make use of the previous version.

The data compilation subsystem is a web interface so that human data editors can pick up the required data from the papers and store them in the server in an appropriate format.
We have made a minor update for the subsystem to include information on the population.

The data registration subsystem is a converter from text-based data obtained by the data compilation subsystem to the data set to be handled by the users of the database.
Two major updates are implemented in this subsystem, which are described in the next subsections.
We also updated the data format along with the update of the data retrieval subsystem as described below.

The data retrieval subsystem is a web interface so that users can access and select data based on various criteria, and then inspect the selected data on a diagram with user-specified axes.
We have made some updates in the subsystem.
First, in the latest version of the subsystem, it automatically produces a text-based data table after the search so that users can download and use it in their local computers.
The data table includes the records of object name, coordinates, reference, metallicity, effective temperature, surface gravity, and all the related values specified by the user.
Other updates are related to the data reprocessing as described below and the details are described in \S~\ref{sec:update}.

\subsection{Update of the database for Galactic halo stars}\label{sec:update}

There are three major updates since the prior version of the database.
We have included the data of some disk stars and metal-rich stars by removing the metallicity threshold of $\feoh = -2.5$.
These metal-rich stars are taken from 14 papers that measure elemental abundances for more than 50 stars in the metallicity range of $\feoh > -2.5$ \citep{Bensby2005,Bensby2014,Takeda2005a,Takeda2005b,Takeda2007,Mishenina2012,Mishenina2013,Reddy2006,AlvesBrito2010,Ruchti2011,Venn2004,Nissen2011,Chen2000}. 
The number of added stars for the metal-rich population is 2380.
The total number of abundance data in these papers is 33727 for 28 element species.

Another update is to improve the consistency of adopted solar abundances among different papers.
As discussed in \citet{Suda2012}, there is a systematic difference in relative abundances scaled by solar values.
In the previous version of the database, the abundances were converted by $\xbyfe = \xbyh - \feoh$ if $\feoh$ values were given in the original papers.
However, this procedure presumed that the adopted solar abundances were common in any papers.
If the abundances of neither \textrm{[X/H]} nor \textrm{[X/Fe]} for element \textrm{X} were available, the abundances were converted from \logeps\ using the solar abundances of \citet{Grevesse1996}.
Therefore, the abundances were not consisntent in any sense when compared with different papers.
On the other hand, in the latest version, we properly take account of the difference in adopted solar abundances during the data registration process to the database server.
If the adopted solar abundances are described in the original paper, all the abundance data in units of \xbyh\ and \xbyfe\ are converted to \logeps.
Otherwise, we use the solar abundance of \citet{Grevesse1996} to convert the abundances, as in Paper I.
The abundances are converted again to \xbyh\ and \xbyfe\ for given sets of solar abundances.
The new version of the database has a choice of the solar abundances such as \citet{Anders1989}, \citet{Grevesse1996}, and \citet{Asplund2009} when users search and plot the data on the data retrieval subsystem (Paper I).

The third major update is the definition of the fiducial values for individual stars based on statistics.
Some data for abundances and stellar parameters have multiple measurement by different authors and/or observational setups.
It is important to fix the fiducial values of these parameters in individual objects to classify the objects such as ``EMP'' stars and ``CEMP'' stars.
We determined the fiducial abundances by (1) taking an average for all the derived data, (2) taking a median for all the derived data, or (3) selecting a representative value according to the priority following the selection criteria described in appendix~\ref{sec:priority}.

We have redefined the class of stars subject to the corrected abundances following the above procedures for all the data.
The classification of the data is the same as in Paper II:
The ``EMP'' population denote the stars with $\feoh \leq -2.5$ according to the drastic change of the evolution of AGB stars \citep{Fujimoto2000,Suda2004,Suda2010} that may characterise the properties of CEMP stars.
Carbon-rich star population labeled ``CEMP'' and ``C-rich'' denote the stars with $\cfe \geq 0.7$ following the discussion in \citet{Aoki2007b} and Paper II.
Giant stars labeled ``RGB'' are defined by the stellar parameters with $\teff \leq 6000$ and $\logg \geq 3.5$.
The stars that do not satisfy this criteria are labeled as ``MS'' in the present version, which could include some horizontal branch stars.

According to these updates, we have extended the choice of datasets in the data retrieval subsystem.
The data registration subsystem produces the abundances of element X in units of $\log \epsilon({\rm X})$, [X/Fe], and [X/H], plus nine sets of stellar classifications according to the choice of reference solar abundances and the criteria for the fiducial abundances and stellar parameters.

In the following, we adopt the fiducial values as abundances normalised by \citet{Asplund2009} with the average of all the reported values.

\subsection{Reassignment of Star Names in the Database}\label{sec:starid}

We have renamed all the stars in the dwarf galaxies registered in the database.
This is to avoid the confusion of star names in individual galaxies because star IDs are sometimes identical in different galaxies and are different for the identical object.
For example, star ``166'' exists in the Leo II and the Sculptor, and have two different names such as ``166'' and ``LeoII-166'' in different papers.
Another example is a star ``UMi297'', ``COS347'', or ``UMiBel10018'', which refers to the same object.

First, we analyse the position and brightness of all the stars registered in the database.
The brightness of the stars are defined by V-band magnitude that are taken from the compiled paper or the VizieR at CDS\footnote{URL: http://vizier.u-strasbg.fr/viz-bin/VizieR}.
The position data are also taken from the literature or the VizieR.
We defined new star IDs by demanding that two stars are within $3$ arcsecs apart, and that their difference in V band magnitude is less than $0.4$ mag.
We also checked the identity of stars by changing the criteria for separation and brightness, and added 49 stars that were confirmed to be identical with their distances larger than $3$ arcsecs, and removed 19 stars that were confirmed to be different objects with their distances within $3$ arcsecs (see appendix~\ref{sec:id}).
Screening of 8017 stars identified 2524 stars that refer to the same object with different IDs.

New star IDs are assigned as {\it SAGA\_[galaxy name]\_[ID]}.
The name of the galaxies are constellation names and abbreviations defined by the International Astronomical Union.
Star IDs are six digit numbers and assigned by the order of publication years and the ascending order of declination within the papers.
All the star IDs defined here are linked to the original names and accessible to the original papers in the same manner as in the SAGA database for Galactic halo stars (see Paper I).
Once the star IDs are assigned, they will not be changed by future updates.
If we decided to compile papers published before 2016, we will not reorder the star IDs, but will allocate an increment to the current ID.
The list of all the star IDs and their original star names is available online and will be kept updated.

\subsection{The data}

Table~\ref{tab:data} summarises the registered data in the database in the descending order of the number of unique stars in each galaxy, except for the bottom one in which only iron abundance data are available so far.
The first and second column shows the name of the dwarf galaxies and the number of registered stars in total, respectively.
The third to 15th column gives the number of unique stars with measured elemental abundances.
The number of data is counted for \feoh\ for iron and \xbyfe\ for other elements.
The last column shows the number of stars analysed with high-resolution spectroscopy where we set $R \geq 15000$ in the data retrieval subsystem.
It is to be noted that the values of $R$ are compiled from the literature by the criterion that the maximum value in the observational setup is taken as a representative of the observations.
In the current version of the database, the high resolution mode of the FLAMES/GIRAFFE on VLT and the MIKE setups satisfy the criterion as well as the Subaru HDS, the UVES on VLT, and the Keck HIRES.
The stars without the information regarding the resolution in the original papers are not counted, except for the cases for which abundances are measured with high resolution spectrographs so that sufficient resolutions must be achieved.

\input{Table1.tex}

\section{Notes on the usage of the combined data}\label{sec:notes}

Users of the database are cautioned in using the combined data from different literature.
Not only are differences in stellar parameters, model atmospheres, solar abundances, spectral resolution important, but there is also an underlying inhomogeneity in the database due to different groups that use different spectral lines.
Discrepancies may be even more pronounced if the authors use different atomic data, hfs corrections, and so on for similar spectral lines.
We expect these errors are random and within $1 \hyp 2 \sigma$ errors, but they can also be systematic.
As stated below, this seems to be the case for Ti between the two Shetrone et al.'s papers.
We only adopt LTE abundances as in Paper I, and hence the homogeneity of the database is not affected by NLTE correction.

In this section, we briefly describe the difference in the method of analysis among papers to get a general grasp of the limitation in using the combined data.
Some comments and cautions in using the data are described below for individual galaxies, in terms of the number of data, method of abundance analysis employed, and adopted solar abundances in the individual papers.
We also present some examples of the difference in abundance data for the same stars in different papers.

\subsection{Carina}

The Carina dSph galaxy is one of the most studied dwarf galaxies in the local group.
On the contrary to rich photometric data ($\sim 1000$) reaching the main sequence, the detailed abundances are available only for $\sim 100$ stars located at the tip of the giant branch. 
The data by \citet{Shetrone2003}, \citet{Koch2008a}, and \citet{Venn2012} are primarily from VLT/UVES with some from Magellan/MIKE, whereas \citet{Lemasle2012} is from VLT/FLAMES, and \citet{Fabrizio2015} only re-analyse the data from the VLT archives.
\citet{Shetrone2003} provided the abundances of 20 elements for five stars in Carina.
\citet{Koch2008a} added 10 stars to the sample, and \citet{Venn2012} added new analyses for 9 stars (4 in common with Koch et al.).
The significant depletions of Na, Mg, Ba, and Zn, as well as other elements, are found in some stars in both studies, e.g., $\xbyfe \le -0.5$ (see \S~\ref{sec:ggc}).
Further 35 stars were added to the sample by \citet{Lemasle2012} from VLT/FLAMES high resolution analyses, and the entire sample of VLT spectra were homogeneously re-analysed by \citet{Fabrizio2015}.

It is to be noted that any two of the literature adopt different solar abundances in converting the abundance units.
Nevertheless, the discrepancy in abundances does not play a role in the discussion of abundance trends in any of the literatures because the largest discrepancy is at most $\sim 0.2$ dex.

There are no overlapping of reported abundances for Carina except for iron abundance.
The difference in \feoh\ is $0.18$ dex on average and $0.49$ at maximum.

\subsection{Fornax}

The abundances of red giants are studied by many papers.
The SAGA database includes 11 papers which report abundances for 1645 stars.
For six papers, adopted solar abundances are not mentioned.
Other papers have a large variety of adopted solar abundances.

\begin{figure*}
  \begin{center}
    \includegraphics[width=13cm]{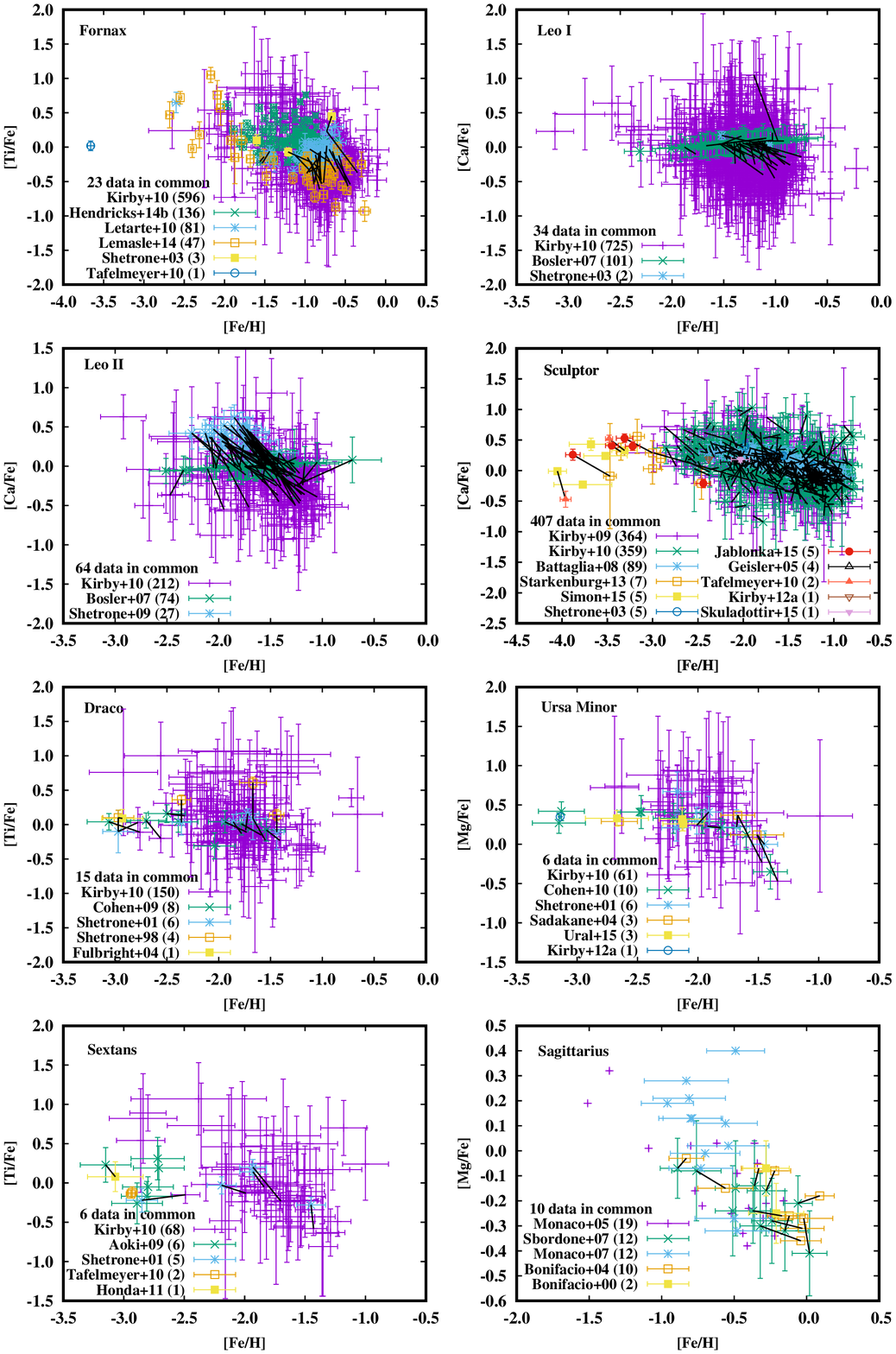}
  \end{center}
  \caption{Abundances of \afe\ as a function of \feoh\ in selected dwarf galaxies whose data are separated by the literature.
      The identical stars are connected by black solid lines to see the abundance difference among various papers.
      Reference with the number of plotted data (in parenthesis) are given in each panel: Kirby+10: \citet{Kirby2010}, Hendricks+14b: \citet{Hendricks2014b}, Letarte+10: \citet{Letarte2010}, Lemasle+14: \citet{Lemasle2014}, Shetrone+03: \citet{Shetrone2003}, and Tafelmeyer+10 : \citet{Tafelmeyer2010}, Bosler+07: \citet{Bosler2007}, Shetrone+03 : \citet{Shetrone2003}, Shetrone+09 : \citet{Shetrone2009}, Kirby+09 : \citet{Kirby2009}, Battaglia+08 : \citet{Battaglia2008}, Starkenburg+13 : \citet{Starkenburg2013}, Simon+15: \citet{Simon2015}, Jablonka+15 : \citet{Jablonka2015}, Geisler+05 : \citet{Geisler2005}, Kirby+12a : \citet{Kirby2012a}, Skuladottir2015 : \citet{Skuladottir2015}, Cohen+09 : \citet{Cohen2009}, Shetrone+01 : \citet{Shetrone2001}, Shetrone+98 : \citet{Shetrone1998}, Fulbright+04 : \citet{Fulbright2004}, Cohen+10 : \citet{Cohen2010}, Sadakane+04 : \citet{Sadakane2004}, Ural+15 : \citet{Ural2015}, Aoki+09 : \citet{Aoki2009}, Honda+11 : \citet{Honda2011}, Monaco+05 : \citet{Monaco2005}, Sbordone+07 : \citet{Sbordone2007}, Monaco+07 : \citet{Monaco2007}, Bonifacio+04 : \citet{Bonifacio2004}, and Bonifacio+00 : \citet{Bonifacio2000}.
      }
  \label{fig:diff}
\end{figure*}

We should be careful about different methods for abundance determinations employed in different papers.
Figure~\ref{fig:diff} shows the contribution of different papers to selected elemental abundances.
The values of [Ti/Fe] versus [Fe/H] are presented for Fornax stars in the top left panel.
There are non-negligible difference among different papers.
A large scatter for [Ti/Fe] and/or [Fe/H] in \citet{Kirby2010} data, as discussed in \S~\ref{sec:knee}, are shown by the black solid lines in the figure.
Among the papers plotted in the figure, the data by \citet{Hendricks2014b} are based on the CaT equivalent widths, different from other papers where the abundances are determined by the spectral fitting with derived stellar atmospheres.
Since there are no stars in common for \citet{Hendricks2014b} with others, we cannot estimate typical differences caused by different method for abundance determinations, but there seems to be a systematic difference in abundances.

\subsection{Leo I and Leo II}

The Leo I and Leo II dSph are distant galaxies among the dSphs registered in the database as shown in figure~\ref{fig:dist} and are discussed in \S~\ref{sec:dist}.
In most cases, the abundances are reported for stars in both galaxies in the same paper.
Most of the data for these populations come from \citet{Kirby2010} (813 stars for Leo I and 258 stars for Leo II) and may be subject to uncertainties associated with the analyses as discussed in \S~\ref{sec:dist} and ~\ref{sec:knee}.
\citet{Kirby2012a} reported Li-rich giants for four stars in Leo I and two stars in Leo II from the sample in \citet{Kirby2010}.
Other studies that produce a number of data \citep{Bosler2007,Gullieuszik2009,Koch2007,Shetrone2009} employ fitting formulae to derive [Fe/H] from CaT equivalent width based on \citet{Carretta1997} or \citet{Zinn1984}.
The abundance data based on high-resolution spectra are available only for 2 stars in Leo I \citep{Shetrone2003}.
The description about the adopted solar abundances is found only in two papers \citep{Kirby2010,Shetrone2003} out of seven papers on Leo I and Leo II.

The plots of [Ca/Fe] versus \feoh\ is shown in figure~\ref{fig:diff} for Leo I and Leo II.
There is a more than 1 dex difference in calcium abundance for SAGA\_LeoI\_000141 (aka LeoI-17226 in \cite{Bosler2007} and LeoI 22692 in \cite{Kirby2010}).
The original star IDs are registered in the SIMBAD database as an identical object.
There is a large disagreement in the reported abundances among \citet{Kirby2010}, \citet{Bosler2007}, and \cite{Shetrone2009} in Leo II, while the discrepancy is relatively small between \citet{Kirby2010} and \citet{Bosler2007}.
It seems that the discrepancy is mainly caused by the estimate of iron abundances because of the slope of close to $-1$ for the majority of the solid lines in the figure.
However, we should be careful that the adopted solar abundances for \citet{Bosler2007} and \citet{Shetrone2009} are not available in the literature, so we assumed that they employed \citet{Grevesse1996}.
This causes $0.17$ dex lower difference in iron abundances than \citet{Anders1989} that \citet{Kirby2010} employed.
This is larger than the average difference of $0.12$ between \citet{Kirby2010} and \citet{Bosler2007}, so the discrepancy is within the uncertainties associated with the adopted solar abundances.
On the other hand, the average difference is $0.47$ between \citet{Kirby2010} and \citet{Shetrone2009}, which is much larger than the potential difference in adopted solar abundance for the case.
For the case between \citet{Bosler2007} and \citet{Shetrone2009}, the average difference is $0.22$, which is relatively small, but the reason for the discrepancy is not clear.

\subsection{Sculptor}

The abundance data are provided by 13 papers, most of which come from the papers by Kirby and his colleagues \citep{Kirby2009,Kirby2010,Kirby2012a,Kirby2012b,Kirby2015}.
The total number of stars (without duplication) is 964.
Other large contribution is provided by \citet{Lardo2016} (94 stars) using the VLT/VIMOS and by \citet{Battaglia2008} (93 stars) using the VLT/FLAMES.
Among them, only \citet{Battaglia2008} provide data based on CaT equivalent widths.
Other studies \citep{Starkenburg2013,Shetrone2003,Simon2015,Geisler2005,Tafelmeyer2010,Skuladottir2015} give minor contribution in statistics but much information on elemental abundances thanks to high resolution spectroscopy.
Adopted solar abundances are provided except \citet{Battaglia2008} and \citet{Kirby2012a}, although the latter uses the same sample as in \citet{Kirby2010} and provides lithium abundances for which the normalisation by solar values is not required except for \feoh.

The abundances of calcium versus metallicity is plotted in figure~\ref{fig:diff}.
As shown by the longest solid line in the figure, the star SAGA\_Scl\_000128 (aka 248 in \cite{Kirby2009} and Scl 1013366 in \cite{Kirby2010}) shows a huge difference in derived metallicity.
\citet{Kirby2009} report the value of \feoh\ to be $-3.01$, while \citet{Kirby2010} derived $\feoh = -1.56$.
The stars are $0.325$ arcsecs apart according to the coordinates in the original paper and referred to as 2MASS J01001452-3347501 in the SIMBAD database.
The discrepancy of the abundances is not likely due to the identification problem because no other stars are registered near the coordinate in SIMBAD.

\subsection{Draco}

Two Kirby's papers \citep{Kirby2010,Kirby2015} provide abundance data for 479 out of 537 stars in the database.
Rich data with high- and low-resolution spectroscopy are available in Shetrone's papers \citep{Shetrone1998,Shetrone2001,Shetrone2013}.
The detailed elemental abundances are also available from the data based on high resolution spectra \citep{Fulbright2004,Cohen2009,Tsujimoto2015}.
Among them, four papers \citep{Kirby2010,Kirby2015,Cohen2009,Tsujimoto2015} provide information on adopted solar abundances, all of which are different each other.

Common data are available for 11 stars for titanium abundances as shown in figure~\ref{fig:diff}.
The largest difference is the case for SAGA\_Dra\_000004 (aka D267 in \cite{Shetrone1998}, Draco 267 in \cite{Shetrone2001}, and Dra 656889 in \cite{Kirby2010}) where the discrepancy is more than $0.5$ dex between \citet{Shetrone1998} and \citet{Shetrone2001}.
\citet{Shetrone2001} reported lower Ti abundance than \citet{Shetrone1998}, which is consistent with the Ti abundance derived by \citet{Kirby2010}.
It is to be noted that both of Shetrone's papers provide a reasonable agreement for other elements such as Na, Mg, Ca, Cr, Fe, Ni, and Ba.
The reason for the discrepancy is likely to be the use of different spectral lines since \citet{Shetrone2001} state that they have added more and better Ti I and Ti II lines.
Apart from this case, there seems to be a good agreement for abundance data in Draco.

\subsection{Ursa Minor}

Ursa Minor is well studied by many papers using the DEIMOS and HIRES on Keck I and the Subaru/HDS.
The number of stars with derived abundances using DEIMOS \citep{Kirby2010,Kirby2015}, HIRES \citep{Cohen2010,Shetrone2001,Ural2015,Kirby2012a}, and HDS \citep{Sadakane2004,Aoki2007} is 299, 20, and 4, respectively.
Adopted solar abundances are available in seven papers out of eight papers in the database.

There are five data in common for magnesium abundances in Ursa Minor.
Among them, Mg abundances for SAGA\_UMi\_000001 (aka UMi 297 in \cite{Shetrone2001}, UMiBel 10018 in \cite{Kirby2010}, and COS 347 in \cite{Sadakane2004}) and SAGA\_UMi\_000006 (aka UMi 199 in \cite{Shetrone2001}, UMiBel 20058 in \cite{Kirby2010}, and COS 82 in \cite{Sadakane2004}) are derived by three independent analysis.
\citet{Kirby2010} give smaller Mg abundance by $\sim 0.6$ dex for these stars, while the other two studies based on high-resolution spectroscopy give consistent results for any elements as well as for Mg.
The discrepancy is not due to iron abundance because their derived iron abundances are consistent within $\sim 0.05$ dex.

\subsection{Sextans}

The majority of the abundance data for Sextans are supplied by \citet{Battaglia2011} and \citet{Kirby2010}.
The two papers produce abundance data for 315 stars, 174 of which are based on the metallicity scale using CaT equivalent width \citep{Battaglia2011}.
The rest of the abundance data are available with high-resolution spectra using the Subaru/HDS \citep[6 and 1 stars, respectively]{Aoki2009,Honda2011}, the Keck/HIRES \citep[5 stars]{Shetrone2001}, and the VLT/UVES \citep[2 stars]{Tafelmeyer2010}.
The adopted solar abundances are not common in any of the papers, except for two papers \citep{Battaglia2011,Shetrone2001} that are not available in the literature.

There are four stars in common for derived Ti abundances in Sextans between \citet{Shetrone2001} and \citet{Kirby2010}.
The difference is not significant for all these stars in $\alpha$-elements commonly reported by both papers.
We also see a minor difference in the abundances of SAGA\_Sex\_000008 (aka S15-19) between \citet{Aoki2009} and \citet{Honda2011}.

\subsection{Sagittarius}

The papers on Sagittarius provide large dataset using high-resolution spectra \citep{Monaco2005,Monaco2007,Bonifacio2000,Bonifacio2004,Sbordone2007,Keller2010}.
Each paper uses different normalisation with solar abundances.
However, two papers \citep{Bonifacio2000,Monaco2007} give no information on the adopted solar abundances.

We have 10 data that enable us to compare Mg abundance derived by two different papers \citep{Bonifacio2004,Sbordone2007}.
All the data with multiple measurement show a reasonable agreement with a typical difference of $0.3$ dex, as shown in the bottom right panel of figure~\ref{fig:diff}.
In addition, there are no systematic difference between any two papers.
This suggests that we can safely compare the abundances of combined data.

\subsection{Other galaxies}

Abundance data other than [Fe/H] with high resolution spectroscopy are available for B\"ootes I \citep{Feltzing2009,Gilmore2013,Ishigaki2014,Norris2010b}, Ursa Major II \citep{Frebel2010}, Hercules \citep{Aden2011,Koch2013,Koch2008b}, Segue 1 \citep{Norris2010c}, B\"ootes II \citep{Koch2014a}, Coma Bernices \citep{Frebel2010}, Leo IV \citep{Simon2010}, Reticulum II \citep{Ji2016a,Ji2016b,Roederer2016}, and Triangulum II \citep{Kirby2017,Venn2017}.
Such data are available for less than 10 stars in each galaxy.

For low-resolution spectra, \citet{Martin2007} provide Fe abundances for B\"ootes I, Canes Venatici I, Ursa Major I and II, and Willman I, based on the scaling relation between [Fe/H] and the equivalent widths of the Ca II lines \citep{Carretta1997}.
The same procedure is employed to derive Fe abundances for B\"ootes II in \citet{Koch2009}.
The abundance analyses with medium resolution ($R = 5,000$) by \citet{Norris2010a} provide many C abundances for B\"ootes I and Segue 1.
The C abundances for B\"ootes I are also provided by \citet{Lai2011}.
The abundances of Segue 2 are only provided by \citet{Kirby2013a}.

Many of the data provide useful information in the sense that adopted solar abundances and stellar parameters are available in the literature.
More data will help to compare the abundance characteristics with those in MW.

\section{Distance to the Individual Stars as a Consistency Check for the Data}\label{sec:dist}

Stars in the local group galaxies with known distance help to constrain stellar parameters, using the comparison of distances estimated from stellar parameters and photometric data.
All the stars in the same galaxy should have the same distances with the negligible amount of uncertainties on the positions in the galaxy, as far as the membership is correct.
In this section, we estimate the distance to stars and check the consistency with the distance to their host galaxies derived from photometric data.

The distance to individual objects and their photometry are related by $V - M_{\rm V} = -5 + 5 \log d$ where $V$ is the apparent magnitude in V band, $M_{\rm V}$ is the absolute magnitude, and $d$ is the distance in parsecs.
The value of $M_{\rm V}$ is estimated by observables using the relation,
\begin{eqnarray}
M_{\rm V} = M_{{\rm bol}, \odot} - {\rm BC} - 2.5 \left( \log \frac{M}{\msun} - \log \frac{g}{g_{\odot}} + 4 \log \frac{T_{\rm eff}}{T_{{\rm eff}, \odot}} \right).
\label{eq:mv}
\end{eqnarray}
Here we assumed that all the stars have $0.8 \msun$, which is consistent with the papers in which surface gravities are determined by the fitting of stellar evolution models.
An exception is the analysis in \citet{Letarte2010} where they adopted $1.2 \msun$ as an initial mass and iterated with derived \logg.
This results in an underestimate of the distance for 108 stars in Fornax by $22$ per cent.
The bolometric corrections, BC, in V band are calculated subject to the prescription in \citet{Alonso1999}.
Therefore, the dependence of distance on the parameters can be expressed as follows.
\begin{eqnarray}
\log d = \frac{1}{5} \left( V + {\rm BC (\feoh, \teff)} \right) + \frac{1}{2} \log M - \frac{1}{2} \logg + 2 \log \teff + {\rm const.}
\label{eq:dist}
\end{eqnarray}
All the available parameters in the equation provide a consistency check among observed values and model parameters.

\begin{figure*}
  \begin{center}
    \includegraphics[width=160mm]{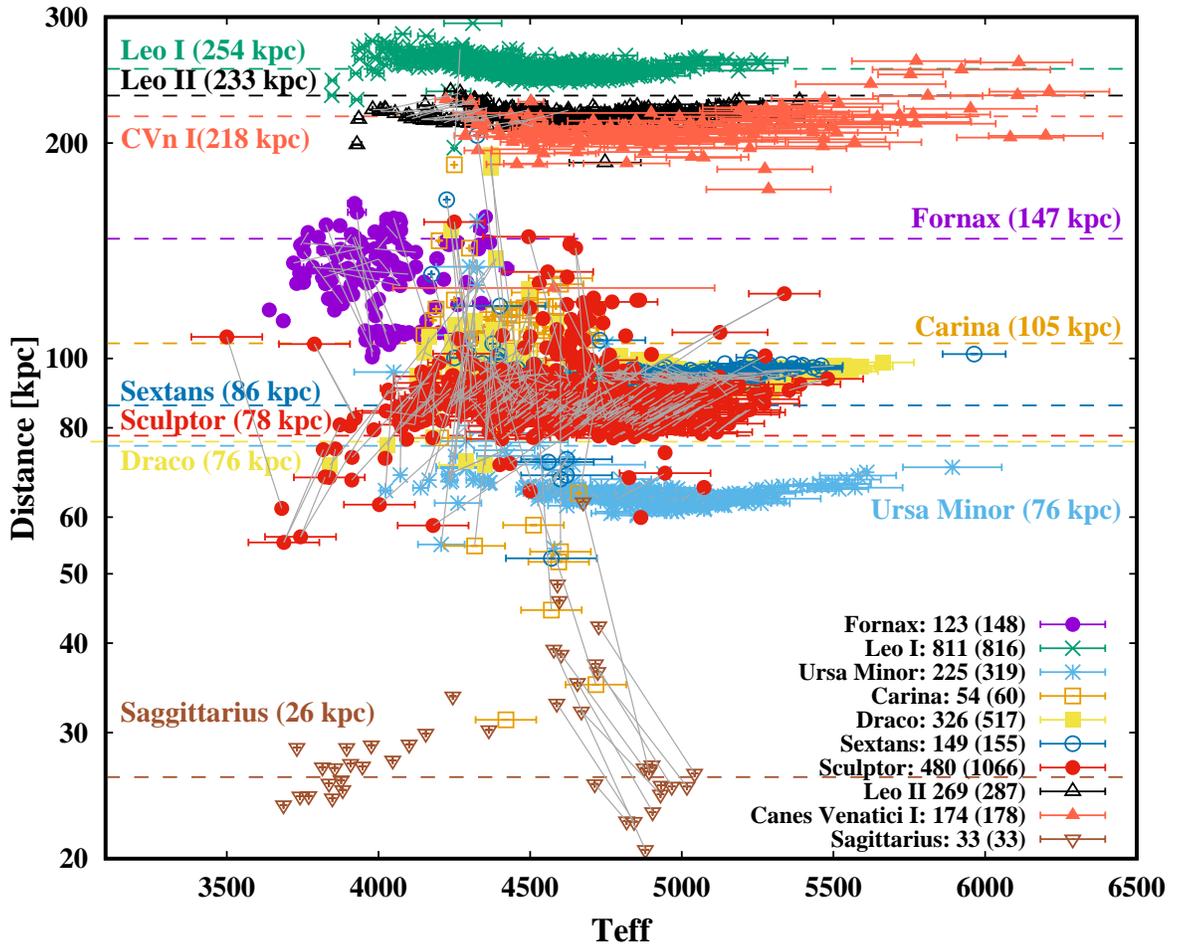}
  \end{center}
  \caption{Distance of individual stars in dwarf galaxies as a function of $T_{\rm eff}$. The distance is calculated according to equation(\ref{eq:dist}). The number of unique stars and the total number of stars (in parenthesis) plotted are given for selected galaxies. The horizontal lines are the distances to galaxies taken from the literature which is given by the labels next to the lines.
  The distances to galaxies are taken from \citet{McConnachie2012}.
  The gray lines connect the same objects in different papers.
      }
  \label{fig:dist}
\end{figure*}

Figure~\ref{fig:dist} shows the distance to stars in selected dwarf galaxies for which all the relevant observables in equation(\ref{eq:dist}) are available.
Photometrically determined distances to individual galaxies are shown by the horizontal lines, taken from the compilation by \citet{McConnachie2012}.
Although the dispersion in the estimated distance is much larger than the size of the galaxies, the deviations are within the uncertainties of stellar parameters in most cases.
In most galaxies, the scatter of estimated distances are within $10$ per cent, which is translated into $\sim 0.08$ dex in \logg.

It may be of concern that there are large systematic differences in galaxies such as Carina, Sextans, Sculptor, Draco, Ursa Minor, and Sagittarius.
Overall underestimate for Fornax can be explained by the different choice of stellar mass in \citet{Letarte2010} as mentioned above.
Most of the stars in Carina have the estimated distance of $\sim 130$ kpc with the minimum of $31.3$ kpc (SAGA\_Car\_000258 aka LG04c\_000626, \cite{Koch2008a}) and the maximum of $190.9$ kpc (SAGA\_Car\_000001 aka CarM10, \cite{Shetrone2003}).
Both extremes are based on the data with high resolution spectra using the UVES on VLT and with the same procedure to determine the stellar parameters.
The scatter is too large to be explained by any uncertainties of observed values.
Some stars in Draco show the large discrepancy of distance for the same object.
The star SAGA\_Dra\_000001 shows various distance estimates: 91.6 or 93.2 kpc (Dra598482, \cite{Kirby2010,Kirby2015}), 184.7 or 192.4 kpc (Draco119, \cite{Shetrone1998,Shetrone2001}), and 83.7 kpc (D119, \cite{Fulbright2004}).
Indeed, these discrepancies come from the difference of adopted surface gravity by $> 0.7$ dex, while the derived elemental abundances are not so different (typically 0.1-0.2 dex and at most 0.4 dex for Na).
Some stars in Sagittarius also show very different distance estimates for the same star.
This comes from the different estimate of \logg\ in \citet{Bonifacio2000} and \citet{Sbordone2007} where the typical difference is $0.4 \hyp 0.5$ dex.

A factor of $\approx 2$ difference in the distance estimate for SAGA\_Dra\_000076 can be alleviated by increasing the derived values of $\log g$ by $\approx 0.6$ dex with the other parameters unchanged.
On the other hand, the same amount of difference requires the change of \teff\ from 4000 K to $\approx$ 5400 K with fixed $\log g$ if we ignore the corresponding change of the bolometric corrections. 
This huge underestimate of \teff\ is very unlikely.
If we try to explain the inconsistency of distances by photometry, $V$ magnitude needs to be decreased by $\approx 1.5$ mag for the fixed $\log g$ and \teff\ of the most evolved giants, which is much larger than the uncertainties associated with the observations.
Therefore, the main cause of the discrepancy should be the overestimate or underestimate of $\log g$.

The estimated distance in Leo I, Leo II, CVn I, Fornax, Sculptor, Draco, and Ursa Minor shows a weak correlation between \teff\ and distance.
The majority of the data come from Kirby and their colleagues \citep{Kirby2010,Kirby2015}.
They employ their own technique to derive atmospheric parameters, which is somewhat different from other studies, and may cause these unexpected correlations.

More work will be needed to understand why the discrepancy happens.
This problem may be related to the crude approximation to derive stellar parameters by the assumption of local thermodynamic equilibrium in one-dimensional atmosphere.

\section{Global Characteristics of Stars in Dwarf Galaxies}

\begin{figure*}
 \begin{center}
      \includegraphics[width=160mm]{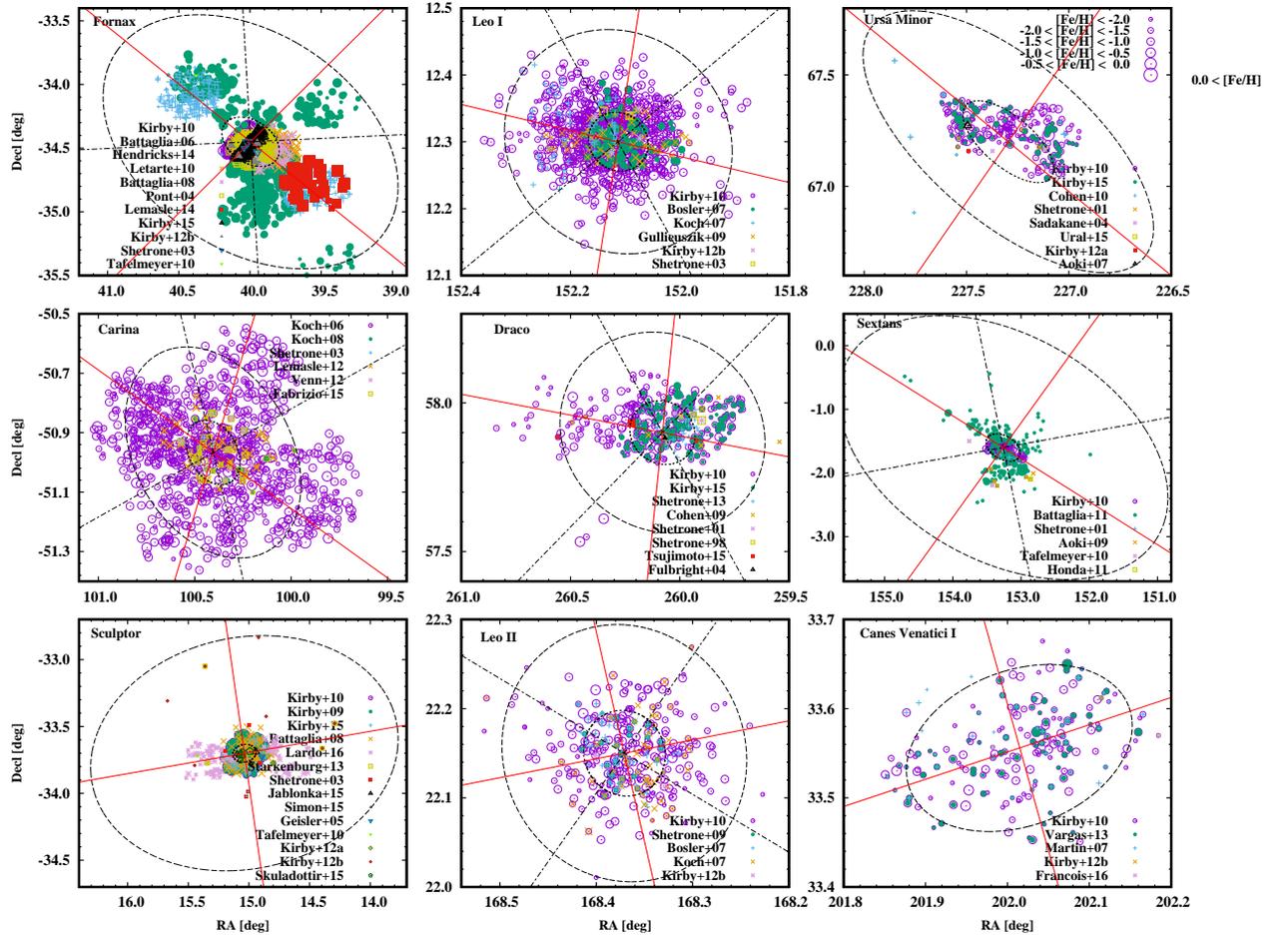}
   \end{center}
  \caption{Spatial distribution of the sample.
  The data are separated by different source as shown by the reference in each panel.
  The size of the symbols denote the metallicity of stars as shown by the legend in the top right panel.
  The two circles represent the King core and tidal radii of the galaxies and are taken from \citet{Mateo1998}, except CVn I where the half light radius is drawn using the value given by \citet{Zucker2006}.
  Two solid lines in each panel are the major and minor axes using the position angle data from \citet{Zucker2006} (CVn I) and \citet{Mateo1998} (others).
  The dashed lines divides the sky coverage to examine the radial distribution (figures~\ref{fig:dir1} and \ref{fig:dir2}).
  The data sources for Fornax are, Kirby+10: \citet{Kirby2010}; Battaglia+06: \citet{Battaglia2006}; Letarte+10: \citet{Letarte2010}; Battaglia+08: \citet{Battaglia2006}; Pont+04: \citet{Pont2004}; Lemasle+14: \citet{Lemasle2014}; Kirby+15: \citet{Kirby2015}; Kirby+12b: \citet{Kirby2012b}; Shetrone+03: \citet{Shetrone2003}; and Tafelmeyer+10: \citet{Tafelmeyer2010}. For Carina, Koch+08: \citet{Koch2008a}; Koch+06: \citet{Koch2006}; Lemasle+12: \citet{Lemasle2012}; Venn+12: \citet{Venn2012}; Farbrizio+15: \citet{Fabrizio2015}. For Sculptor, Kirby+09: \citet{Kirby2009}; Starkenburg+13: \citet{Starkenburg2013}; Geisler+05: \citet{Geisler2005}; Kirby+12a: \citet{Kirby2012a}; Skuladottir+15: \citet{Skuladottir2015}. For Leo I, Bosler+07: \citet{Bosler2007}; Koch+07: \citet{Koch2007}, Gullieuszik+09: \citet{Gullieuszik2009}. For Draco, Shetrone+13: \citet{Shetrone2013}; Cohen+09: \citet{Cohen2009}; Shetrone+01: \citet{Shetrone2001}; Shetrone+98: \citet{Shetrone1998}; Tsujimoto+15: \citet{Tsujimoto2015}; Fulbright+04: \citet{Fulbright2004}. For Leo II, Shetrone+09: \citet{Shetrone2009}; Koch+07: \citet{Koch2007}. For Ursa Minor, Cohen+10: \citet{Cohen2010}; Sadakane+04: \citet{Sadakane2004}; Aoki+07: \citet{Aoki2007}; Ural+15: \citet{Ural2015}. For Sextans, Battaglia+11: \citet{Battaglia2011}; Aoki+09: \citet{Aoki2009}; Honda+11: \citet{Honda2011}. For Canes Venatici I, Martin+07: \citet{Martin2007}.
    }\label{fig:coord}
\end{figure*}

The spatial distribution of the sample for selected galaxies are shown in figure~\ref{fig:coord}.
The data are grouped by references to show clear differences in sky coverages among different studies.
The stellar identification procedure as discussed in \S~\ref{sec:starid} is not considered in the figure, i.e., some stars are overplotted in the same coordinate.

In the following discussion, we should keep in mind that the sample is neither complete nor homogeneous.
In most galaxies, combined observations do not cover all the area within the tidal radius.
In addition, the spatial distribution is not symmetric in any direction.
There should always be a sampling effect for the analysis below, which is difficult to correct.

The figures in this paper are created using the data retrieval subsystem with the following options:
only data sets from the same papers are chosen if two or more papers report elemental abundances.
The priority parameters are applied to data in figures~\ref{fig:mdf}, \ref{fig:cum}, \ref{fig:cemp}, \ref{fig:cfe}, \ref{fig:mgfe}, \ref{fig:mgslope}, and \ref{fig:slopedsph}.
The number of plotted stars in each galaxy is displayed in the figures if relevant.
The numbers in parenthesis are the total number of data in the case that the data for the same star are available in multiple papers.

\subsection{Metallicity distribution function}
\begin{figure*}
 \begin{center}
      \includegraphics[width=160mm]{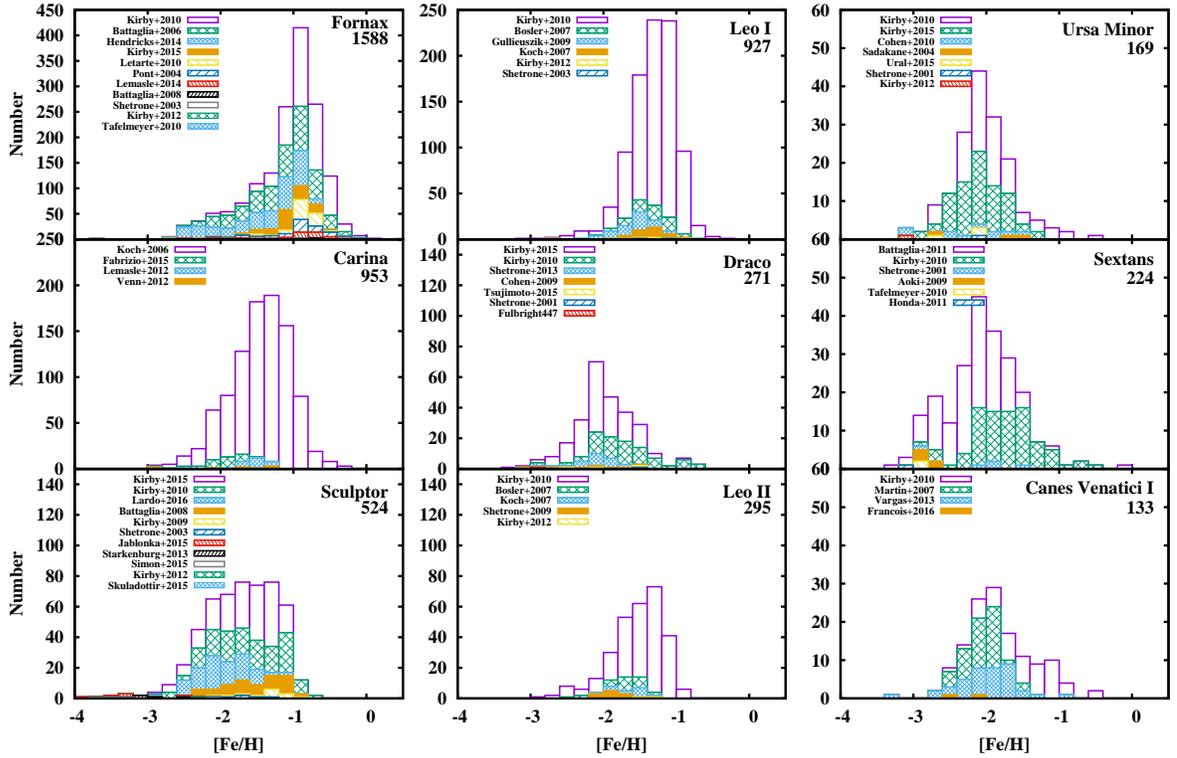}
   \end{center}
  \caption{Metallicity distribution function for each galaxy. The data are separated by different source as denoted by the reference in the top left corner. The galaxy name and the number of data are shown in the top right corner. If iron abundances are reported by multiple papers, the star is counted in a paper with the highest priority according to our criteria (see appendix~\ref{sec:priority}).
    }\label{fig:mdf}
\end{figure*}

Figure~\ref{fig:mdf} shows the metallicity distribution of each galaxy that has more than 100 data on $\feoh$.
The data are compiled from all the available literature irrespective of the quality of the spectra, but we excluded stars with uncertainties greater than $0.3$ dex, following \citet{Kirby2011a}.
The dominant contributors to the metallicity distributions are the data with low-resolution spectroscopy due to the efficiency of data acquisition using multi-fiber spectrographs on Keck or VLT.

The shape of the MDF has some variations from galaxy to galaxy, although the combinations of different sources may cause biases.
As shown in figure~\ref{fig:coord}, a sampling bias depends on the target selection in individual galaxies and on the radial gradient as discussed below.
Selecting stars in the outer parts of the dwarf galaxies will bias toward lower metallicity, but this is not the case for any galaxies from the inspection of figure~\ref{fig:coord}.
More importantly, we can expect that all the data are randomly sampled or, in most cases, limited by the brightness of stars, in any of the compiled papers.
Therefore, the selection bias in metallicity and other elemental abundances should be minimal.
This is different from the case of the Galactic halo stars, where the authors of the papers usually focus more on extremely metal-poor stars rather than normal Population I and II stars.

\begin{figure*}
 \begin{center}
      \includegraphics[width=160mm]{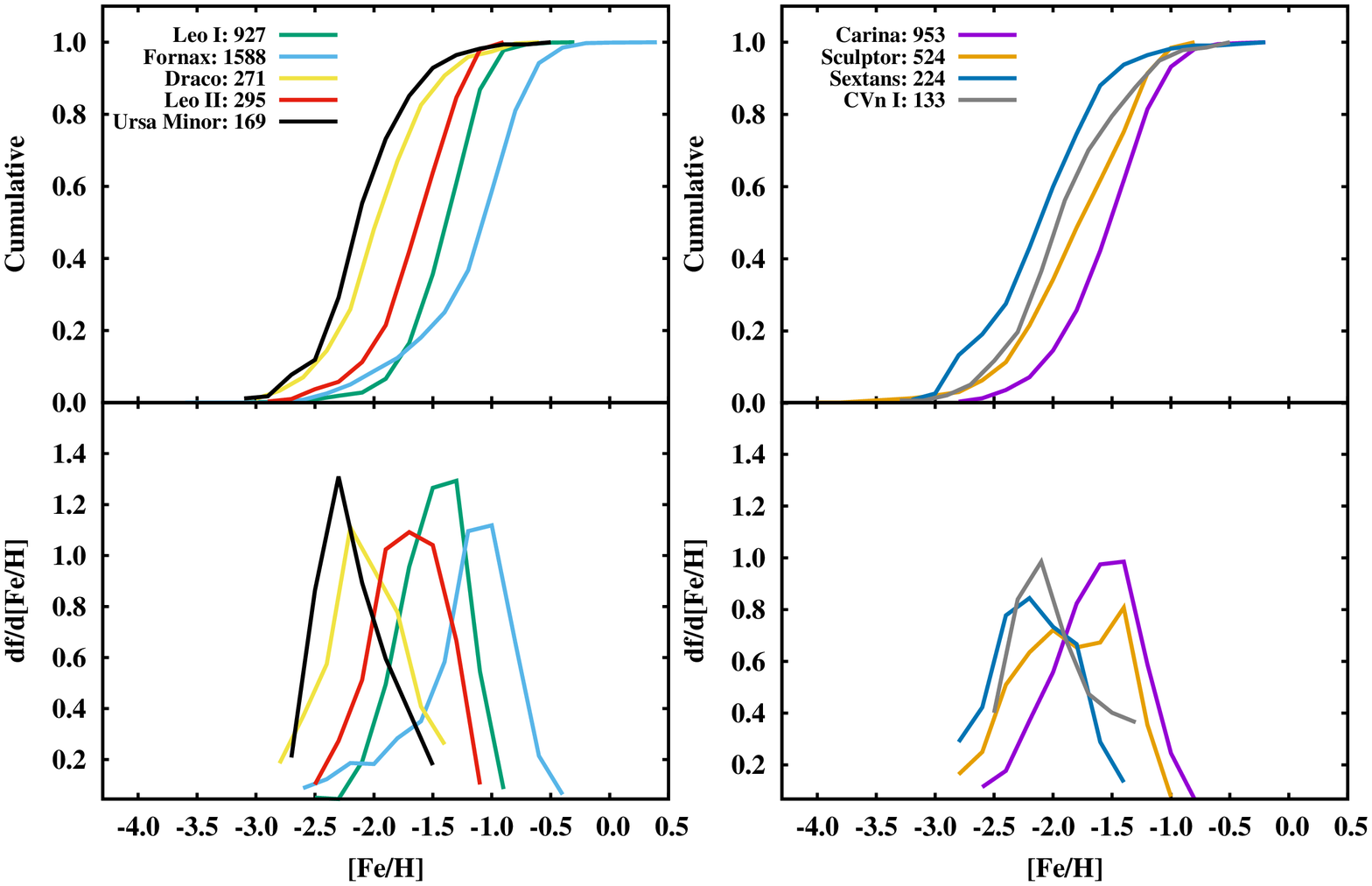}
   \end{center}
  \caption{Top: cumulative metallicity distribution for individual galaxies. The sample is the same as in figure~\ref{fig:mdf}. Bottom: slope of the cumulative MDF. See text for details.}
  \label{fig:cum}
\end{figure*}

A different feature can be found in the slope of the MDF at the left and right side of the peak.
The left side of the MDF represents the activity of star formation and metal-production from massive stars.
The right side probably reflects the quenching of star formation and the saturation of metallicity in the interstellar medium.
Apparently there is a dichotomy of the slope.
The sharp increase in the distribution may correspond to an initial star burst, while the gradual increase may reflect an extended star formation history.
This can be quantified by cumulative metallicity distribution as shown in the top panels of figure~\ref{fig:cum}.

The bottom panels show the change of the slope with respect to metallicity, which is calculated by $df / d\feoh |_{i} = ( f_{i+1} - f_{i} ) / (\feoh_{i+1} - \feoh_{i} )$, where $f_{i}$ is the number of stars in the cumulative MDF at $i$-th bin.
The bin width for metallicity is set at $0.2$ dex.
We exclude the bins in which the number of data is less than five.
Note that, theoretically, the derivatives of the cumulative MDF gives a MDF.
The advantage of using this analyses is that we can normalise the distribution with the total number of observed stars, not with the peak height of the MDF.
Our analysis helps to compare the MDF of different populations.
Clearly, the distribution of the slope is separated into two groups according to its maximum, as shown in the left and right panels.
This is also the case even if we change the size of the bin width from $0.1$ to $0.4$ dex.
Leo I, Fornax, Draco, Leo II, and Ursa Minor (left) show steeper increase in the MDF before the peak compared with Carina, Sculptor, CVn I, and Sextans (right).
The former may represent an initial star burst episode due to the efficient production of metals in the early phase of chemical evolution.
On the other hand, the case of continuous star formation results in a less steep slope because the increasing rate of star formation produce relatively fewer metal-rich stars.

Interestingly, our simple interpretation of the observations reasonably agrees with the star formation history inferred from the stellar population of dwarf galaxies using the color-magnitude diagram (e.g., \cite{deBoer2012a,deBoer2012b,Weisz2014,VandenBerg2015}).
According to the results of \citet{Weisz2014} and \citet{deBoer2012a}, Draco and Ursa Minor belong to the galaxies that experienced initial star burst, while Carina, Sculptor, Leo I, and Fornax are thought to experience extended star formation.
\citet{Weisz2014} show that the star formation history of Leo II is in the middle of both, i.e., star formation activity continues until $\sim 6$ Gyr.
They also show that the model for CVn I suggests two major star formation episodes.
The star formation history of Fornax is continuous, but has a peak at $\feoh \sim -1$ \citep{deBoer2012b}, which may correspond to the steep rise of MDF at the same metallicity range.
Apart from the complex cases such as CVn I and Leo II, only Leo I apparently contradicts to our predictions.

The dependence of the shape of MDF on star formation history is not simple.
The mechanism to make a slope of the MDF has complex aspects and is mainly controlled by star formation rate or the timescale of star formation which is determined by central density, total mass, and so on.
Depending on the star formation timescale, the trend can be opposite, i.e., initial star burst results in a less steep MDF and continuous star formation results in a steep slope.
Star formation history also depends on inflow and outflow of gas in the system.
The details of the parameter dependence can be found in the simulations of \citet{Hirai2015}.
Similar kind of discussions are found for the low-metallicity tail of the MDF at $\feoh < -2.5$ \citep{Starkenburg2010,Karlsson2012}, although the sampling bias is inevitable at this low metallicity range.

The shape of the metal-rich tail in the MDF may be of interest as well.
Figure~\ref{fig:mdf} shows clear cutoffs for Sculptor, Leo I, and Leo II at the right side of the peak of the MDF.
This can be seen in the bottom panels of figure~\ref{fig:cum}, where the sharp drops of the slope are observed.
These qualitative differences may be biased by uncertainties of the determinations of metallicities with low-resolution spectroscopy as well as the combination of data from multiple papers.
More data with homogeneous spectral analyses and spatial coverage will characterise the star formation history of the dwarf galaxies from the slopes of the cumulative MDF. 

Apparent double peaks in the MDF of Sextans are contributed by \citet{Battaglia2011} for the lower peak and by \citet{Kirby2010} for the higher peak.
The lower peak is argued to be the result of the contamination by star cluster members \citep{Battaglia2011,Karlsson2012}.
In particular, \citet{Karlsson2012} insist that one of the stars in the lower peak show the anti-correlation of O and Na, which is typical in the galactic clusters, by their analysis of the result of \citet{Aoki2009}.
On the other hand, \citet{Battaglia2011} detected a difference in the kinematics of stars in the lower peak.
A small peak is also found in Ursa Minor at $\feoh \simeq -3$.
We cannot conclude if the stars in the peak are the relics of globular clusters because we do not have the data of proton-capture elements for the population.
See more discussion in \S~\ref{sec:ggc}.

\citet{Kirby2011a} pointed out the existence of bumps at $\feoh \sim -2.5$ for Draco and Leo II, while they are not found in figure~\ref{fig:mdf}.
The reason for the absence of bumps in our data is the choice of bin width of $0.2$ dex in the figure, considering the uncertainties associated with the abundance analyses in individual data and the potential biases caused by combining the data from different literature.
When we set the bin width at $0.1$ dex, small bumps are visible in the same manner as in \citet{Kirby2011a}.
We suspect that the bumps in Draco and Leo II are the result of sampling since about 30 \% of the sample in Draco have uncertainties greater than $0.2$ dex and the majority of stars in Leo II have uncertainties greater than $0.1$ dex. 

\begin{figure*}
 \begin{center}
      \includegraphics[width=160mm]{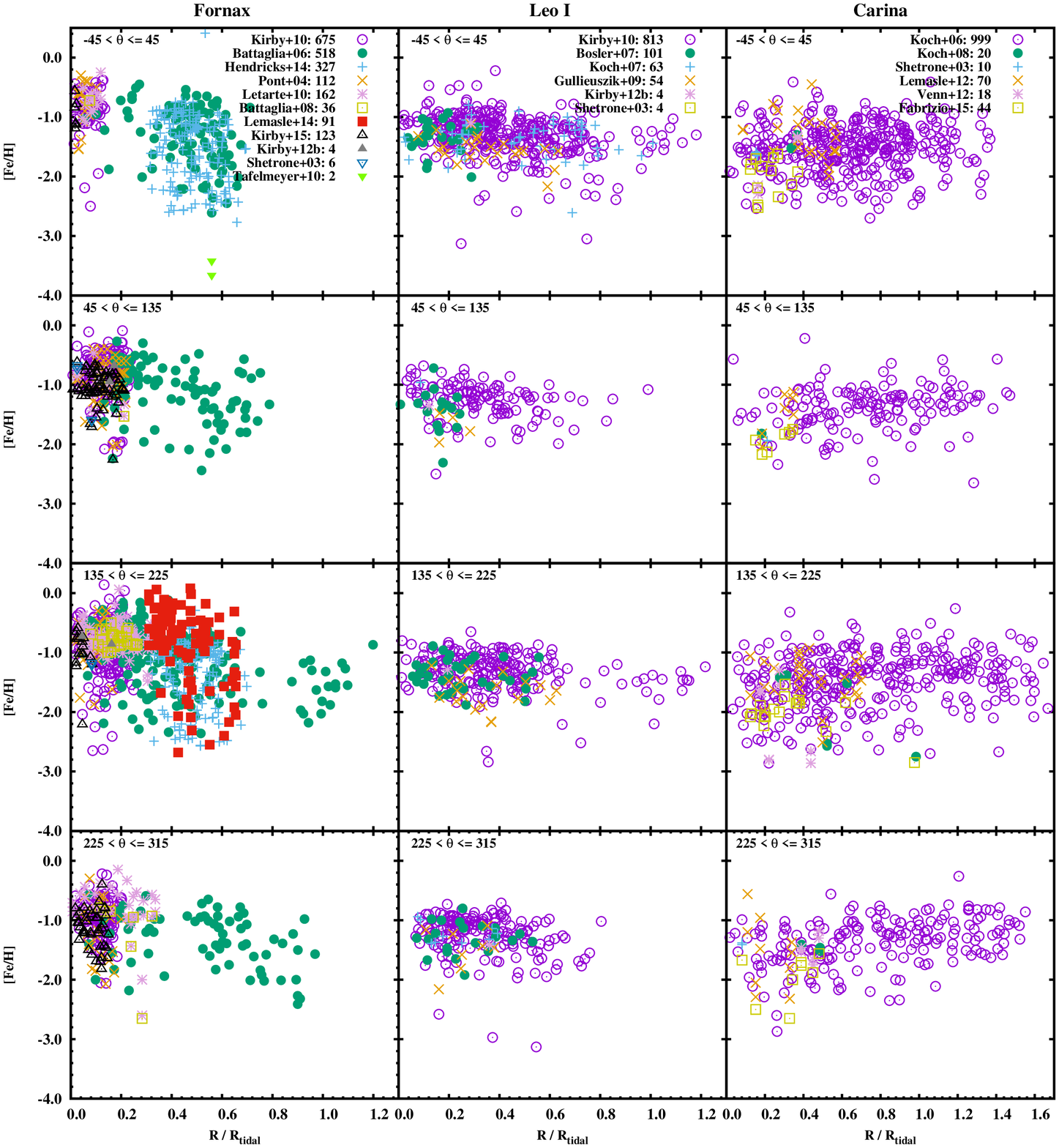}
   \end{center}
  \caption{Metallicity distribution with radial distance for the members of Fornax, Leo I, and Carina in the database. The radial distance is normalised by the tidal radius at the direction from the center. The meaning of the data points are the same as in figure~\ref{fig:coord}. The radial distribution is separated into four directions with $\theta = 0^{\circ}$ set at the NE axis as shown by the solid lines in figure~\ref{fig:coord}. The data selection criteria are the same as in figure~\ref{fig:mdf}.
    }\label{fig:dir1}
\end{figure*}

\subsection{Radial gradient}
\begin{figure*}
 \begin{center}
      \includegraphics[width=160mm]{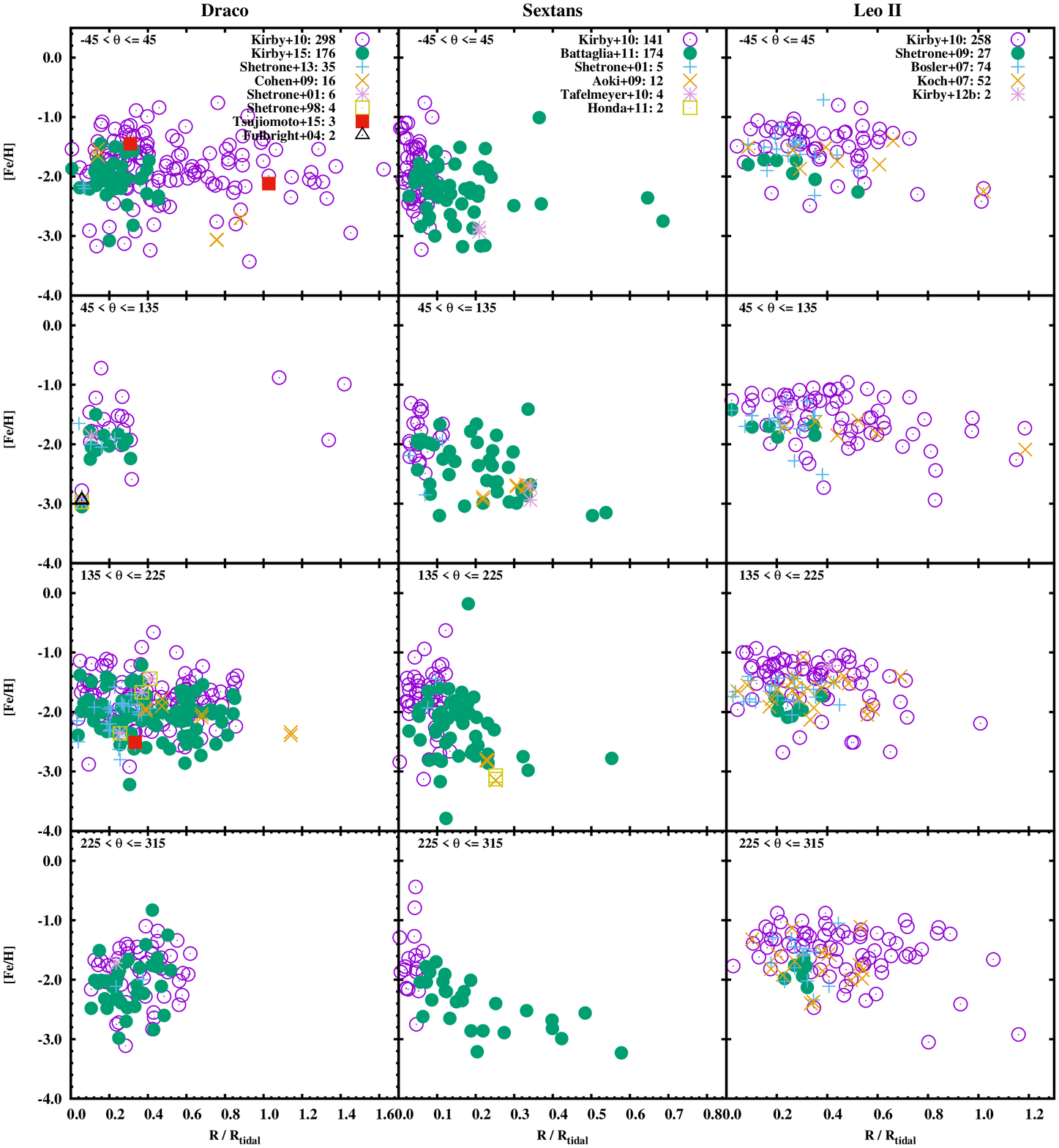}
   \end{center}
  \caption{The same as figure~\ref{fig:dir1}, but for the members of Draco, Sextans, and Leo II.
    }\label{fig:dir2}
\end{figure*}

Figures~\ref{fig:dir1} and \ref{fig:dir2} shows the radial gradient of metallicity with direction dependence for six galaxies.
The sky coverage is divided into four domains along the major axes.
A radially decreasing trend was first reported by \citet{Battaglia2006} for Fornax.
We show in figure~\ref{fig:dir1} for the first time that the trend can be found in any directions.
The combined data do not seem to destroy the trend of the radial gradient, although the systematic difference must be considered to make a reliable discussion, which is beyond the scope of this paper.
We do not see any radial gradient for Carina in any directions, which improves on the lack of an overall gradient reported by \citet{Koch2006}.
Stars in Draco may have a radial gradient in [Fe/H] in NE direction along the major axis, where the data points are dominated by \citet{Kirby2010}.
We checked the gradients of [X/Fe], for example, [Ca/Fe], and do not find any gradients in any direction, which means that the yields from progenitor stars do not alter the abundance ratios.
The metallicity gradient is also observed in Leo I and Leo II in all the directions.
There are several stars with $\feoh \sim -3$ in the outer region of Sextans, which apparently makes gradients.
Again, all the plots may not be statistically meaningful.
Future studies should check if the radial gradient and direction dependence exist or not.

\subsection{Fraction of carbon-enhanced stars}\label{sec:cemp}

The fraction of carbon-enhanced stars in extremely metal-poor (EMP) stars plays a key role to understand the origin of the most iron-poor stars in MW.
It is established that the contribution of carbon-enhanced metal-poor (CEMP) stars to the EMP population is increasing with decreasing metallicity in the Galactic halo \citep{Norris1997a,Lucatello2006,Suda2011,Yong2013,Lee2014}.
Thanks to the large survey of EMP stars \citep{Beers1992,Christlieb2002} together with the observations of CEMP stars with high-resolution spectroscopy \citep{Aoki2007b,Spite2005,Cohen2006}, the number of known CEMP stars is more than 100 if we use the definition of CEMP stars by $\cfe \geq 0.7$ (Paper II, see also \cite{Aoki2007b}).
CEMP stars are divided into subclasses such as \cemps\ and \cempno\ with and without the enhancement of \sit-process elements \citep{Aoki2002a}.
The origins of carbon-enhancement for these subclasses are thought to be different by the difference in binary frequency \citep{Starkenburg2014,Hansen2016a,Hansen2016b}.

\begin{figure}
 \begin{center}
      \includegraphics[width=80mm]{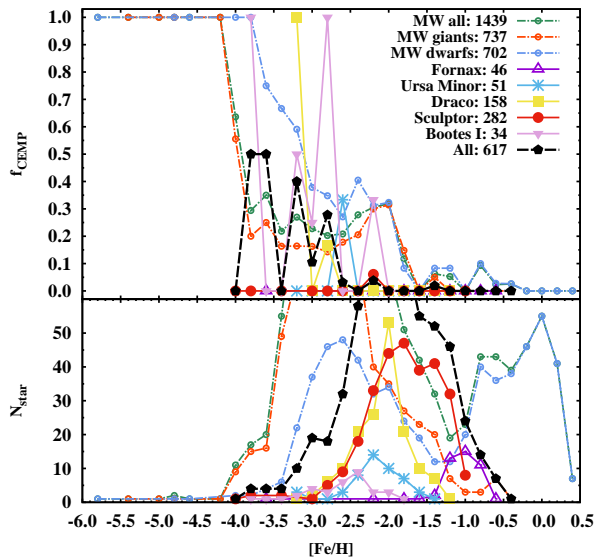}
   \end{center}
  \caption{Fraction of carbon-enhanced stars ($\cfe \geq 0.7$) as a function of metallicity for selected galaxies (top). The numbers of stars with carbon abundance measured are given in the bottom panel. The numbers in parenthesis are the total number of data with carbon abundance measured, but excluding the data with upper limits. The dashed line with circles denotes the fraction of CEMP stars for the combined data of dwarf galaxies in the database, which includes 586 stars from five galaxies in the panel and additional 24 stars from B\"ootes II, Come Ber, Segue 1, Sextans, and Ursa Major II. The CEMP fractions for MW are given for giant stars, main-sequence stars, and both.
    }\label{fig:cemp}
\end{figure}

The fraction of CEMP stars are smaller in dwarf galaxies than MW as shown in figure~\ref{fig:cemp}.
In particular, \cemps\ stars, which is dominated at $\feoh \sim -2$ in MW, is very rare among the sufficient number of sample for Draco and Sculptor.
It is the characteristics of dwarf galaxies that the large enhancement of carbon abundance (typically $\cfe \gtrsim 2$) is not found.
A possible exception may be B\"ootes I.
A similar trend of CEMP fraction can be found in B\"ootes I compared with MW where a large CEMP fraction is found at $\feoh \sim -2.5$.
More data is needed to confirm this since the number of stars are still small at this metallicity range.
For the lowest metallicity range of $\feoh \lesssim -3$, the fraction of CEMP stars (possibly dominated by \cempno\ stars as in MW) is apparently common with MW, although the sample is small.
Currently, all the stars with $\feoh < -3$ have $\cfe < 1.0$, which means that CEMP fraction becomes zero if we define CEMP stars by $\cfe \geq 1.0$. 
This may be in sharp contrast with MW in which the carbon enhancement has a dichotomy \citep[Yamada et al., in prep.]{Bonifacio2015,Yoon2016}.
For the case of the MW halo, the large fraction of CEMP stars requires the change of the initial mass function (IMF) to produce large fraction of AGB binaries at $\feoh \lesssim -2$ \citep{Suda2013,Lee2014,Yamada2013}.
It would be interesting to explore if the high mass IMF peaked at $\sim 10 \msun$ \citep{Komiya2007} should be applied to dwarf galaxies.

\begin{figure}
 \begin{center}
      \includegraphics[width=80mm]{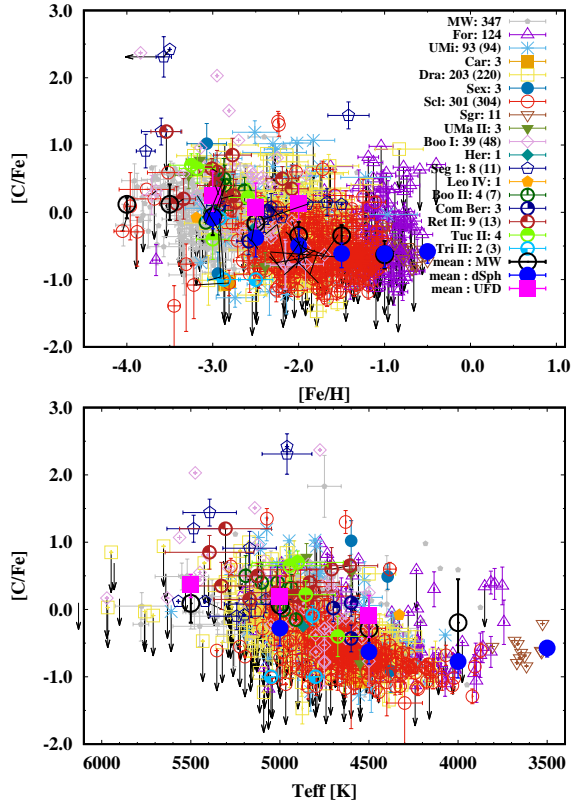}
   \end{center}
  \caption{Top: Carbon abundance as a function of metallicity for MW and the local group galaxies. Only C-normal giants are plotted for MW.. The figures in the parenthesis are the number of plotted data in each galaxy. The large circle represents the average value of [C/Fe] within the 0.5 dex bin of metallicity for MW stars (black open circles) with $\log g \leq 2.0$ and the stars in classical dwarf spheroidal galaxies: Sgr, Car, Dra, For, Scl, Sex, and UMi (blue filled circles) and ultra-faint dwarf galaxies: others (magenta filled squares), respectively. Carbon-enhanced stars ($\cfe \geq 0.7$) are excluded for the means. The error bars denote the 25 and 75 percentile of the value of [C/Fe]. Bottom: Carbon abundance as a function of effective temperature.
    }\label{fig:cfe}
\end{figure}

The top panel of figure~\ref{fig:cfe} shows carbon abundances as a function of metallicity for red giants in MW and dwarf galaxies.
The average carbon abundances are denoted by the large circles.
The data for MW stars are taken for C-normal giants with $\log g \leq 2.0$ to inspect the effect of C-depletion in evolved giants.
Unfortunately, there are only two stars with $\feoh \gtrsim -1$ that have measured carbon abundances, and therefore, we cannot compare with stars in dwarf galaxies.
We separate the average carbon abundances for classical dSphs and UFDs.
\citet{Salvadori2015} argue that the deficiency of the fraction of CEMP-no stars in dSphs is due to the skew in the observed MDF away from very low metallicities, where CEMP-no stars are found.
It is clear that the carbon abundances in classical dSphs are smaller than those in MW at any metallicity range.
It is also evident that the carbon abundances in UFDs are larger than those in MW.
The amount of difference ranges from $0.2$ to $0.4$ dex at $-2.5 \lesssim \feoh \lesssim -1.5$.
The most probable interpretation for this discrepancy is the difference in initial abundances.
One possibility is the contribution of Type Ia supernovae (SNe), which lowers \cfe\ at lower \feoh\ in dwarf galaxies, as discussed in \S~\ref{sec:knee}.
However, we should be careful about the possibility of systematic difference in deriving the carbon abundances.
In any case, the lower fraction of CEMP stars should not be the result of extra mixing, unless the efficiency of extra mixing has a variation in MW and dwarf galaxies.
The reason for or the existence of the discrepancy is to be explored.

Interesting features can be seen in the trend of average carbon abundances.
The increasing carbon abundance with decreasing metallicity is clearly found for both MW and dwarf galaxies.
This is the confirmation of our discussion in Paper II that the average carbon abundances are different for stars above and below $\feoh \sim -2$.
This may support the transition of the IMF as discussed in Paper II, or can be interpreted by the hypothesis that carbon is provided by external sources with its amount almost independent of metallicity, or that the carbon to iron ratio has a metallicity dependence in progenitor stars.

We may need to consider the potential change of carbon abundances in dwarf galaxies where the majority of stars in the database belong to the bright red giant branch and may subject to an evolutionary effect.
The surface abundances of carbon and nitrogen show a decreasing trend with increasing the luminosity (or decreasing the effective temperature) of giants \citep{Gratton2000}.
This is found in extremely metal-poor stars without the large enhancement of carbon \citep{Spite2005}, while it is not for CEMP stars (Paper II).
Possible corrections for carbon abundances are proposed for observed extremely metal-poor stars \citep{Aoki2007b,Placco2014}.
The decreasing trend of \cfe\ with decreasing \teff\ is also observed in dwarf galaxies (bottom panel in figure~\ref{fig:cfe}).
The depletion of carbon in red giants may suggest an extra mixing, which dredged-up the matter in the hydrogen-burning shell where the CN cycles are operating.
In fact, the fraction of CEMP stars can be different between dwarfs and giants in MW (figure~\ref{fig:cemp}).
However, in the case of MW, this is probably due to the detection limit for carbon abundances in dwarfs at low metallicity, where it is difficult to identify carbon-normal turn-off stars.
Since the discrepancy is not observed at $\feoh \gtrsim -2.5$, this observational bias should not be important for dwarf galaxies as most of the stars have $\feoh > -2.5$.

\subsection{Signature of different chemical enrichment of $\alpha$-elements}\label{sec:knee}

The $\alpha$-elements like Mg, Si, Ca, and Ti provides a good measure of the chemical enrichment history of galaxies.
They are synthesised mainly in massive stars and distributed throughout a galaxy during the core-collapse supernova (Type II SN) explosion.
In contrast, Type Ia SNe are not thought to be responsible for $\alpha$-elements while they primarily contribute to the chemical enrichment of iron-group elements.
Therefore, the combination of these two processes in very different timescales determines the average value of \afe.
In other words, the value of \afe\ provides a trace for the fractional contribution of Type II and Type Ia SNe, although there are some arguments regarding the contribution of Type Ia SNe in dwarf galaxies (see e.g., \cite{Nomoto2013}).
Because of the longer timescale of Type Ia SNe than that of Type II SNe, the value of \afe\ is expected to decrease at certain age or metallicity, which is clearly seen in the abundance trend in our Galaxy.
The drop of \afe\ in the \afe\ vs. \feoh\ diagram is called the ``knee'' \citep{Tinsley1979,Matteucci1990}.

\begin{figure*}
 \begin{center}
      \includegraphics[width=160mm]{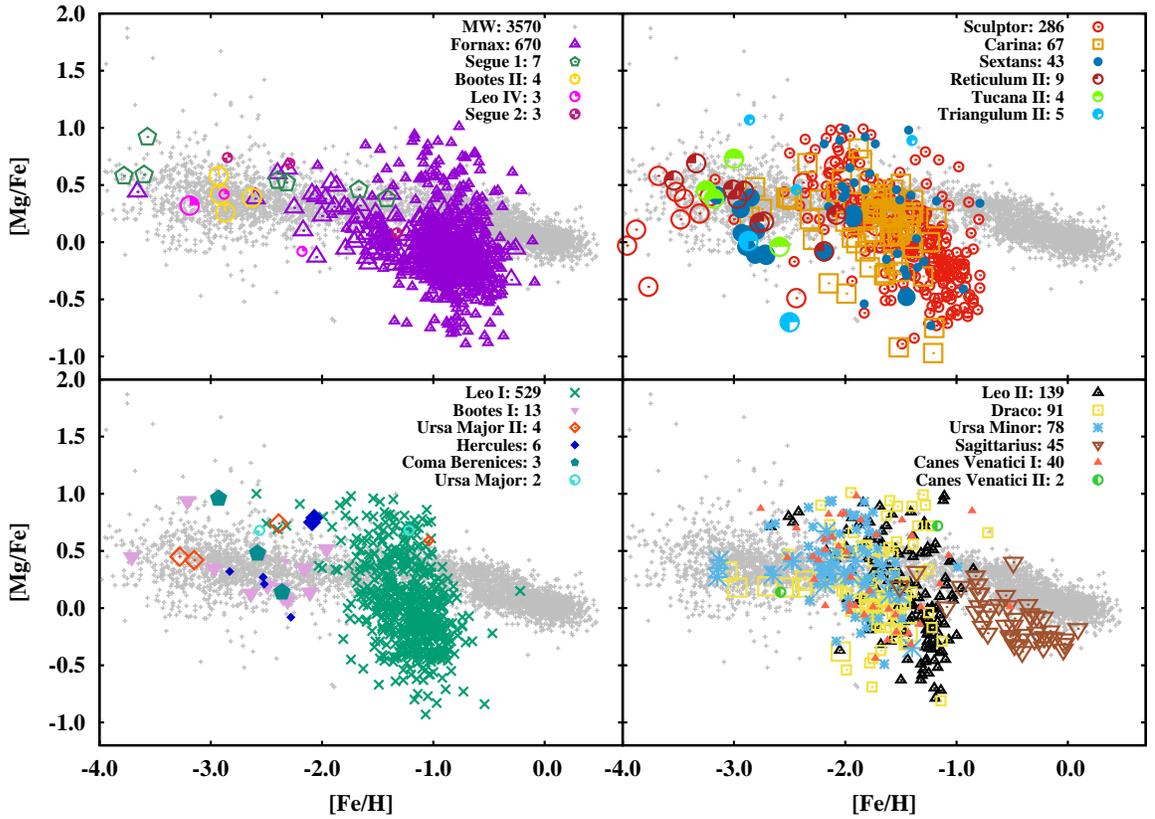}
   \end{center}
  \caption{\mgfe\ as a function of \feoh\ for 23 galaxies in the database. The abundance data based on high-resolution spectra are denoted by large symbols. Stars in MW, taken from the latest version of the database, are also plotted for comparison. The number of plotted data are given next to the name of the galaxy. The data selection for plotting is subject to the defined priority to plot one data for one star. The data with upper limits are excluded from the plot. The error bars are removed for visibility.
    }\label{fig:mgfe}
\end{figure*}

The different position of the knee and the slope of decreasing \afe\ against \feoh\ are easily identified by eye thanks to the large samples in several dSphs (figure~\ref{fig:mgfe}, see also \cite{Tolstoy2009}).
Several studies have tried to identify the position of the knee for individual galaxies and found a difference among galaxies \citep{Cohen2009,Cohen2010,Hendricks2014a}.
Another important difference is that \afe\ is lower than MW in some dSphs.
At least a couple of stars show $\mgfe < 0$ based on high-resolution spectroscopy ($R \geq 15,000$) for Fornax \citep{Letarte2010,Lemasle2014,Hendricks2014a}, Carina \citep{Lemasle2012,Venn2012,Fabrizio2015}, Sculptor \citep{Geisler2005}, Draco \citep{Shetrone1998,Shetrone2001,Cohen2009}, Sextans \citep{Shetrone2001,Aoki2009}, and Sagittarius \citep{Bonifacio2000,Bonifacio2004,Sbordone2007} (see figure~\ref{fig:mgfe}).
In figure~\ref{fig:mgfe}, $50$ stars in Fornax, $11$ stars in Carina, $3$ stars in Sculptor,$2$ stars in Draco, $1$ star in Sextans, and $23$ stars in Sagittarius show $\mgfe \leq -0.2$.
The number of stars are not necessarily large in all the galaxies due to the limited available sample, but this is in sharp contrast to MW by the fact that only $24$ stars have $\mgfe \leq -0.2$ among more than $3000$ MW stars with Mg abundance available in the SAGA database.
These observational facts are studied by theoretical models of chemical evolution (see, e.g., \cite{Lanfranchi2003}), but the mechanism to produce the difference among galaxies is not yet understood.

\begin{figure}
 \begin{center}
      \includegraphics[width=80mm]{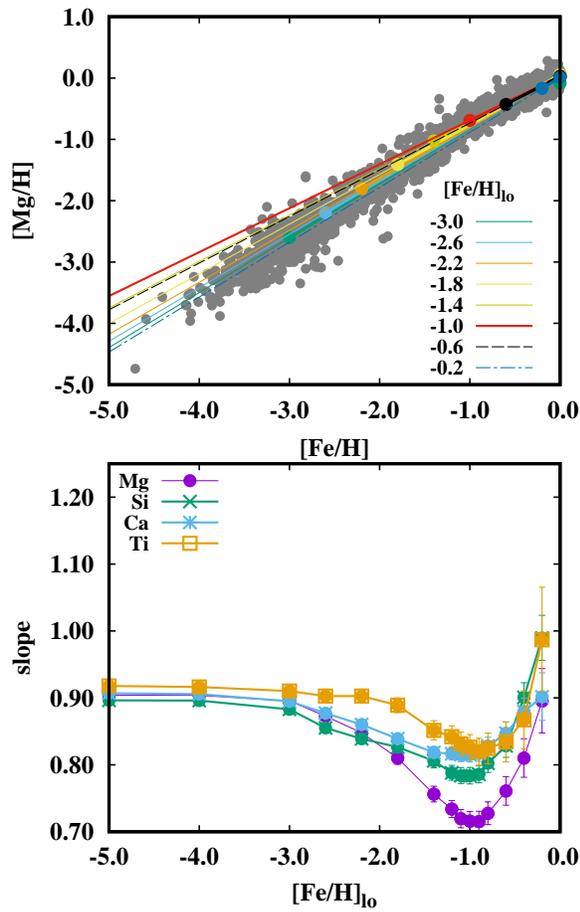}
   \end{center}
  \caption{Top: The abundance ratio \mgh\ versus \feoh\ for MW in the SAGA database, overplotted with the RMA regression lines with different cutoff metallicity $\feoh_{\rm lo}$ (see text). The ranges of fittings are denoted by the thick lines and points. Bottom: The slopes of the RMA regressions for Mg, Si, Ca, and Ti as a function of cutoff metallicity. The error bars show the confidential interval of the slope with 2.5 and 97.5 percentile.
    }\label{fig:mgslope}
\end{figure}

The top panel of figure~\ref{fig:mgslope} shows the fitting to the observed data on the \mgh\ vs. \feoh\ diagram for MW using the reduced major axis (RMA) regression.
We chose the relation between \mgh\ for y-axis rather than \mgfe\ to avoid the uncertainties caused by the estimate of \feoh.
We use the {\it lmodel2} package of {\it R}\footnote{http://www.r-project.org} to analyse the data.
To identify the position of the knee, the RMA regressions are repeated with varying the sample size by setting the lower boundary for metallicity.
We defined a cutoff metallicity, $\feoh_{\rm lo}$, above which the RMA regression is applied.
The slopes obtained by this analysis for four $\alpha$-elements are shown in the bottom panel of figure~\ref{fig:mgslope}.
The position of the knee is identified by the minimum of the slope \footnote{In the ideal case of the contribution of Type Ia SNe, the slope should keep the same value after the minimum. The increase of the slopes after the minimum is not due to the astrophysical reasons, but is due to the truncation of the data in vertical direction with $\feoh_{\rm lo}$, which skews the distribution of data toward lower [X/H] near $\feoh \simeq \feoh_{\rm lo}$. The skewness of the distribution is more noticeable in smaller sample by increasing $\feoh_{\rm lo}$, which results in the larger values of slopes as shown in the figure.}.
Excellent agreement for the position of the knee at $\feoh = -1.0 \pm 0.1$ is obtained.
The slope of the abundance data at $\feoh = -1.0$ may represent how Type Ia SNe contributed to the chemical enrichment.
We define its value as $a$ by setting $\afe = a \feoh + b$ (fitting the data with $\feoh \geq -1.0$).
The value for Mg, Si, Ca, and Ti is $-0.37^{+0.02}_{-0.02}$, $-0.26^{+0.01}_{-0.01}$, $-0.22^{+0.01}_{-0.01}$, and $-0.22^{+0.02}_{-0.02}$, respectively, where the uncertainties correspond to 95 \% confidence interval.

It is to be noted that our sample include all the Galactic component such as thin disk, thick disk, and halo.
\citet{Cohen2010} derive the knee metallicity at -0.53 and -0.50 for thick disk and thin disk components, respectively.
This suggests that our sample is dominated by halo component, although the halo component can be divided into two populations from their abundance patterns \citep{Nissen2010}.
We should think about the possibility of the absence of the knee because it is not necessarily conspicuous from the chemical enrichment history.
As discussed below, the knee position for Draco is not clearly determined.

\begin{figure*}
 \begin{center}
      \includegraphics[width=160mm]{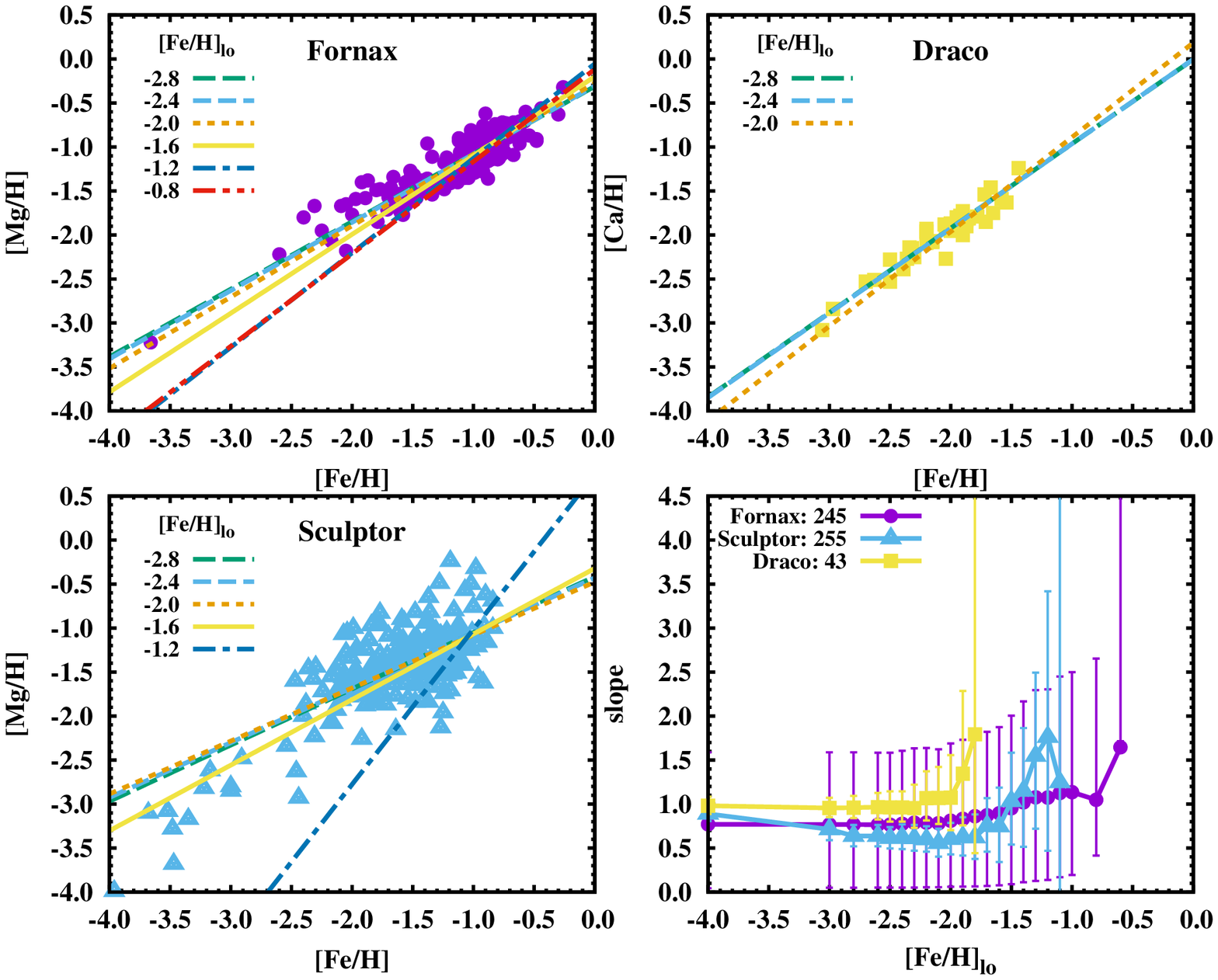}
   \end{center}
  \caption{The same as figure~\ref{fig:mgslope}, but for five selected dSphs for which more than 90 data are available. The bottom right panel shows the slope of the RMA regression for Mg abundances for dSphs. The numbers in the parenthesis next to the galaxy names stand for the total numbers of the sample. Note that [Ca/H] is used for Droco, while [Mg/H] is used for Fornax and Sculptor.
    }\label{fig:slopedsph}
\end{figure*}

Figure~\ref{fig:slopedsph} shows the results of the same analysis applied to the dSph data.
The change of the slope with different $\feoh_{\rm lo}$ does not show clear minimum for dSph data for Fornax and Draco.
This is due to the large uncertainties in Mg (and Ca) abundance or the limited range of metallicity available for the analysis.
The number of data is small for Draco, which is less reliable when the RMA fit is applied to the data with a large cutoff metallicity.
It is important to Increase the number of data for metal-rich stars in order to precisely determine the change of the slope in this analysis.

Table~\ref{tab:slope} provides the position of the knee and its slope for our Galaxy and dSphs for which sufficient number of data are available, together with the literature values. 
All the dSphs have the knee metallicities lower than MW, which implies that the iron-enrichment in dSphs are less efficient than MW before Type Ia SNe play a role.
This can be interpreted as follows: (1) The ejecta of Type II SNe are not retained in the host galaxy due to smaller potential well in dwarf galaxies, and/or
(2) The star formation rate is smaller compared with MW.
On the contrary to the expectation of this interpretation, the metallicity of the knee position is not well correlated with the total mass of the galaxy.
This may be due to the limitation for the reliability of statistical analyses based on medium-resolution spectroscopy.

We have removed the data by \citet{Kirby2010} for Fornax, Sculptor, and Draco since they cause large scatters of abundances and make the data analyses unreliable.
We have omitted the results for Leo I and Leo II due to their poor fitting and results.
Most of the data in Leo I and Leo II are provided by \citet{Kirby2010}, resulting in the slope with $> 1$, which is at odds with observations because [Mg/Fe] should have a negative correlation with [Fe/H].

The comparison of the knee position with previous work is possible for Fornax, Sculptor, and Draco, which is summarised in tab.~\ref{tab:slope}.
For Fornax, our result is consistent with \citet{Tolstoy2009} and \citet{Hendricks2014a}.
\citet{Tolstoy2009} concluded that the knee should be below $\feoh = -1.5$ with their smaller sample than ours.
\citet{Hendricks2014a} derived the knee position at $\feoh = -1.9$ from the comparison of their toy model with their observational sample which consists of their own data for Mg, Si, and Ti abundances based on pipeline procedure and the data of \citet{Letarte2010} based on high-resolution spectroscopy.
The data of \citet{Hendricks2014a} are not published, and therefore they are not included in the SAGA database.
The knee position for Sculptor is estimated to be $\feoh = -1.8$ by \citet{Tolstoy2009}, which reasonably agrees with our result.
The lowest knee metallicity we obtained for Draco is in reasonable agreement with the value derived by \citet{Cohen2009} who estimate the knee position at $\feoh = -2.9$.
We found another minimum at $\feoh \sim -2.8$, although it is marginal.
Due to the insignificant detection of the minimum, there may be a possibility of the absence of the knee for Draco.

\input{Table2.tex}

\subsection{Abundance anti-correlations of O-Na and Mg-Al}\label{sec:ggc}

Dwarf galaxies and the Galactic globular clusters (GGCs) have similar brightness and are sometimes considered to be good targets for comparison, although the properties of these systems are quite different in the content of dark matter, size, star formation history, and chemical abundance homogeneity.
In particular, $\omega$ Cen is thought to be a tidally disrupted dwarf galaxy in the viewpoint of kinematics \citep{Majewski2000,Bekki2003}.
The similarity and difference of abundance patterns in globular clusters are investigated in other systems such as Large Magellanic Cloud \citep{Johnson2006,Mucciarelli2009}, Fornax \citep{Letarte2010}, and M31 \citep{Sakari2016}.
Therefore it is useful to check the connection between MW and dwarf galaxies in the context of stellar abundances in those systems, although there are no guarantees that we are able to compare the stars in GGCs and dSphs in a reasonable manner due to the variations of iron abundances in dwarf galaxies.

\begin{figure}
 \begin{center}
      \includegraphics[width=80mm]{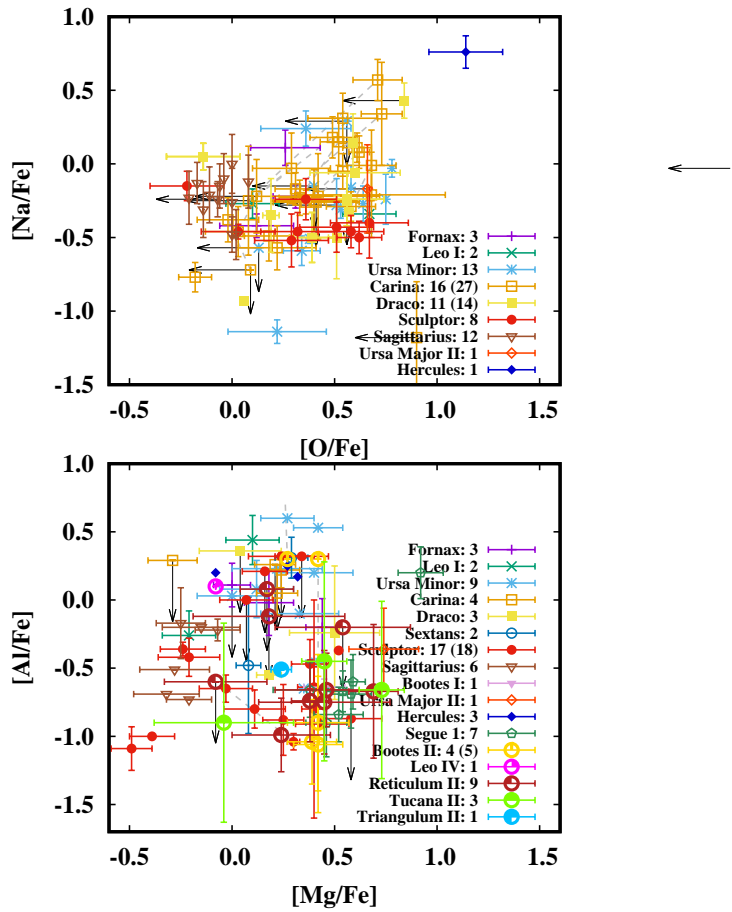}
   \end{center}
  \caption{Correlation of sodium and oxygen abundances (top) and of magnesium and aluminium (bottom). The data with meaningless upper limits are removed. The number in parenthesis is the number of stars plotted in the diagram for each galaxy.
    }\label{fig:ona}
\end{figure}

Figure~\ref{fig:ona} shows the correlation of the abundances of proton-capture elements.
The anti-correlations of the abundances among these elements are established for stars in GGCs (see e.g., \cite{Carretta2009}) including $\omega$ Cen \citep{Norris1995,Johnson2010}, while they are not found in field halo stars (Paper II).
The global trends of these relationships seem to have positive correlations as in the case of metal-poor halo stars (see figures 23 and 24 of Paper II).
Interesting difference is the large scatter of sodium abundances, which spans more than two dex.
The most sodium depleted stars have values of [Na/Fe] $< -1$, in SAGA\_UMi\_000098 \citep[aka UMiCOS171]{Cohen2010} and in SAGA\_Car\_001209 \citep[aka Car-5070]{Venn2012}.
It is in contrast with SAGA\_Her\_000002 \citep[aka Her-2]{Koch2008b}, which shows an enhancement of [Na/Fe] $\sim 1$.

When we inspect the data in individual galaxies, some of them show no positive correlations as well as no clear anti-correlations.
In Sculptor, O, Mg, and Al abundance show $\sim 1$ dex difference, while Na abundance changes within $0.5$ dex without any enhancement ([Na/Fe] $< 0$).
There are eight stars that appear in both panels of figure~\ref{fig:ona}.
Stars showing large Al enhancement, SAGA\_Scl\_000004, \citep[aka SclH479]{Shetrone2003}, SAGA\_Scl\_000003 \citep[aka SclH461]{Shetrone2003} SAGA\_Scl\_000002 \citep[aka SclH459]{Shetrone2003} also show large O enhancement.
On the other hand, O-depleted star, SAGA\_Scl\_000007, \citep[aka 982]{Geisler2005} shows the depletion of Al.
Therefore, there are no connections to proton-capture nucleosynthesis for the apparent Mg depletion and Al enhancement in the bottom panel.
These imply that these stars are unlikely to be associated with abundance anti-correlations.
No anti-correlations in Sculptor was also reported by \citet{Geisler2005} who added four stars and inspected the anti-correlation of O and Na abundances.
Draco also does not support the abundance anti-correlations for both O-Na and Mg-Al, while there are $\sim$ 1 dex variations in O, Na, Mg, and Al abundances.
There is one star, SAGA\_Dra\_000009, that have reported abundances for the four elements \citep[aka 3157]{Cohen2009}.
The abundances of the star ([O/Fe] $= 0.59$, [Na/Fe] $= 0.14$, [Mg/Fe] $= 0.18$, and [Al/Fe] $= -0.55$) do not show any evidence of proton-capture nucleosynthesis.

Sagittarius dwarf spheroidal galaxy is an interesting system due to the strong effect of tidal stripping by MW.
It is revealed that the Galactic globular clusters like Terzan 7 and Palomar 12 share the similar abundance patterns with the Sagittarius dwarf galaxy and that these globular clusters are associated with the Sagittarius stream \citep{Cohen2004b,Sbordone2007}.
This argument is supported by the abundance patterns of Sagittarius stars in figure~\ref{fig:ona} where all the elements are depleted compared to MW and other dwarf galaxies.

Although the current sample is small, it is not likely that stars in dwarf galaxies share the same abundance characteristics with those in the GGCs.
The large variations in the abundances may come from independent source of chemical enrichment such as Type II SNe, Type Ia SNe, and mass loss from AGB stars.

\subsection{neutron-capture elements}\label{sec:ncap}

\begin{figure*}
 \begin{center}
      \includegraphics[width=160mm]{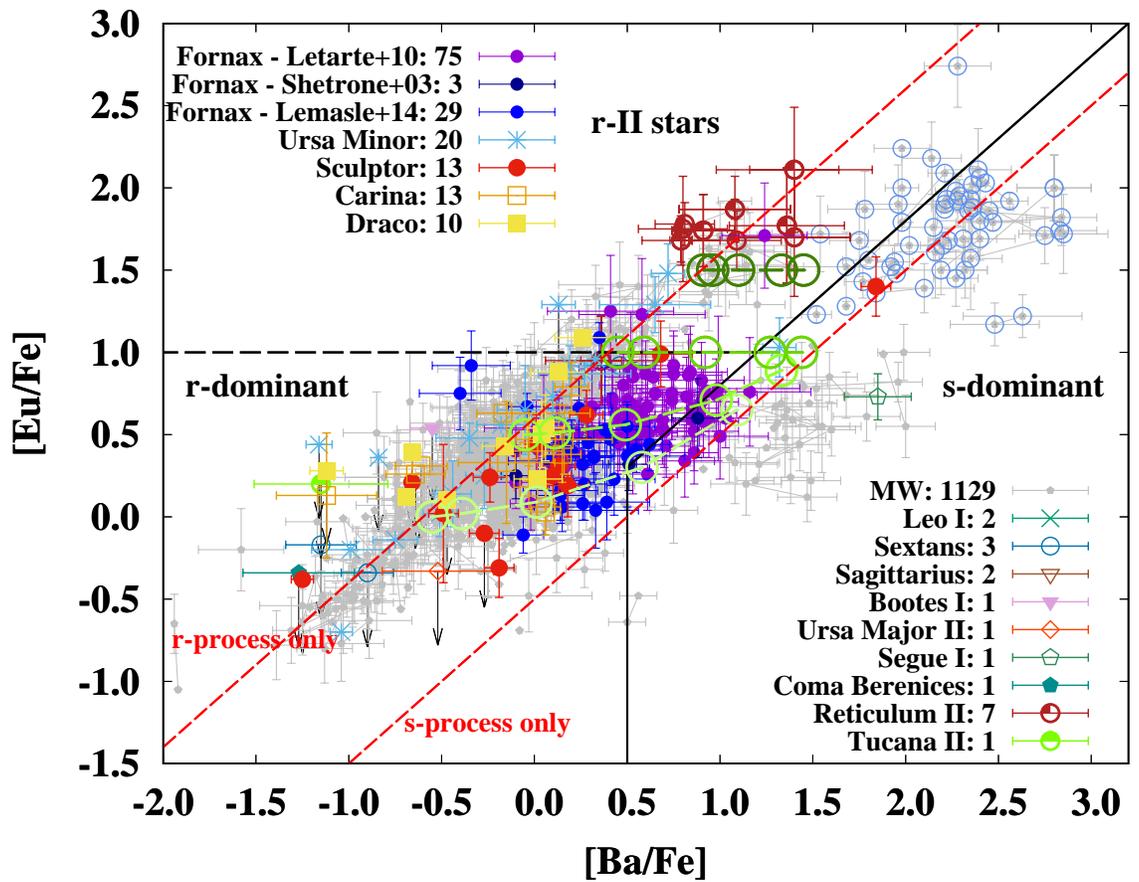}
   \end{center}
  \caption{Distribution of dSph stars on the [Ba/Fe] and [Eu/Fe] diagram. The production site of heavy elements is separated by the solid line of [Ba/Fe] $\geq 0.5$ and [Eu/Ba] $\leq -0.2$ to define {\it s}-dominant population, following Paper II. The horizontal dashed line divides the r-II stars among {\it r}-dominant stars defined as [Eu/Fe] $\geq 1.0$. The data for Fornax stars are separated by references.
    Two dashed lines with the labels ``r-process only'' and ``s-process only'' represent the canonical value of the \rit-[\sit-]process nucleosynthetic results of [Eu/Ba] $= 0.5$ and $-0.7$, respectively.
    The big open circles connected with dashed lines are simple models of the mixture of \rit-process dominated interstellar medium with \sit-process yields, starting with different Eu abundances. see text for details.
    }\label{fig:euba}
\end{figure*}

Figure~\ref{fig:euba} shows the distribution of stars on the [Ba/Fe] and [Eu/Fe] diagram.
As discussed in Paper II, the diagram enables to see a clear separation for the production site of heavy elements, whether the source of neutron-capture elements is ascribed to the \rit-process or the \sit-process.
According to Paper II, we use the criteria, [Ba/Fe] $\geq 0.5$ and [Eu/Ba] $< -0.2$, as shown by the solid line, for stars for which the abundances of neutron-capture elements are dominated by the \sit-process, and call them `\sit-dominant'.
Otherwise, we classify the stars as `\rit-dominant' for which neutron-capture elements should mainly come from the \rit-process.

In contrast to the successful classification for the source of neutron-capture elements for Galactic halo stars (figure~5 in Paper II), the sample stars in dwarf galaxies are not well separated by the solid line in figure~\ref{fig:euba}.
As shown in the figure, most of the sample in the database are supposed to be \rit-dominant from the diagram, except for SAGA\_Seg1\_000004 \citep[aka SDSS J100714+160154]{Frebel2014} where it is clearly located in \sit-dominant region and shows the enhancement of lead abundance, according to the abundances of neutron-capture elements reported by \citet{Frebel2014}.
This is consistent with the fact that the number of \cemps\ stars is small in the sample of dwarf galaxies, as discussed in \S~\ref{sec:cemp}.

However, there are striking difference between the MW stars and Fornax stars.
We clearly see Fornax stars around the boundary of the \rit/\sit-dominant region with the enhanced Ba and Eu abundance.
The abundances of Ba and Eu in Fornax stars are measured by three independent observations using the UVES \citep{Shetrone2003} and the FLAMES/GIRAFFE \citep{Letarte2010,Lemasle2014} on VLT.
Apparently, there is a systematic difference in the data between \citet{Letarte2010} and \citet{Lemasle2014}, although some data in \citet{Lemasle2014} and two out of three stars in \citet{Shetrone2003} belong to anomalous population in which both of [Ba/Fe] and [Eu/Fe] are enhanced.
The majority of this population have $\feoh \sim -1$, i.e., the chemical enrichment with neutron-capture elements proceeded earlier than MW.

To explore the possible origin of this anomalous population, we make a simple model of chemical enrichment.
Let us suppose that the interstellar medium (ISM) in Fornax consists only of \rit-process ejecta for neutron-capture elements below $\feoh = -1$.
From the observed value as shown in the dashed line labeled with ``r-process only'', the mass ratio, $f_{\rm ISM} = \mismeu / \mismba$ is considered to be a known value, where $\mismeu$ and $\mismba$ is the mass of Eu and Ba in the ISM, respectively.
Then, we assume that AGB ejecta having pure \sit-process abundance pattern mix with the ISM.
In the same manner, the ratio of Eu to Ba in ejecta is defined as $f_{\rm ej} = \mejeu / \mejba$, which can be calculated from the mean value of [Eu/Ba] $= -0.5$ for \cemps\ stars in MW.
Now we define the mixing parameter as
\begin{equation}
F_{\rm mix} \equiv \frac{M_{\rm Eu, ej}}{M_{\rm Eu, ISM}} = \frac{f_{\rm ej} M_{\rm Ba, ej}}{f_{\rm ISM} M_{\rm Ba, ISM}}.
\end{equation}
The final value of [Eu/Ba] for given $F_{\rm mix}$ can be written as
\begin{eqnarray}
\euba &=& \log \frac{\mejeu + \mismeu}{\mejba + \mismba} \frac{\xbasun}{\xeusun} \\
&=& \euba_{r} + \euba_{s} + \log \left( F_{\rm mix} + 1 \right) - \log \left( 10^{\euba_{r}} F_{\rm mix} + 10^{\euba_{s}} \right)
\end{eqnarray}
where $\euba_{r}$ is the initial value of [Eu/Ba] $= 0.6$ in the ISM, taken from the mean value for carbon-normal EMP stars in MW, and $\euba_{s}$ the value of [Eu/Ba] $= -0.5$ in pure \sit-process ejecta.
The mass fraction, $\xbasun$ and $\xeusun$ is the solar value for Ba and Eu, respectively.
From this equation, the abundance ratio of [Eu/Ba] in Fornax stars is given as a function of $F_{\rm mix}$, the mixing fraction of the ISM matter and AGB ejecta.

The enrichment of Eu by AGB ejecta can be written as
\begin{equation}
\eufe = \eufe_{r} + \log \frac{F_{\rm mix} + 1}{\frac{\mejfe}{\mismfe} + 1} \label{eq:eu}
\end{equation}
with
\begin{equation}
\frac{\mejfe}{\mismfe} = 10^{\eufe_{r}} \frac{\xeusun}{\xfesun} \frac{\mejfe}{\mejeu} F_{\rm mix}
\end{equation}
where $\mejfe$ and $\mismfe$ is the mass of iron in the AGB ejecta and the ISM, respectively, and $\xfesun$ the solar iron abundance.
Here the ratio $\mejfe / \mejeu$ corresponds to AGB yield and can be estimated from the observed value.
\begin{equation}
\frac{\mejeu}{\mejfe} = 10^{\euba_{s} + \bah_{s} - [{\rm Fe}/{\rm H}]_{\rm ISM}} \frac{\xeusun}{\xfesun}
\end{equation}
where $\bah_{s}$ is the Ba abundance in AGB ejecta and $\feoh_{\rm ISM}$ the metallicity of the ISM during the enrichment process.
We take the value of $\bah_{s} = 0.5$, which is assumed to be the typical abundance in \cemps\ stars around $\feoh_{\rm ISM} = -1$.
Now we obtain the Eu abundance from equation(\ref{eq:eu}) and Ba abundance from $\bafe = \eufe - \euba$ for the given initial Eu abundance, $\eufe_{r}$ and the mixing parameter, $F_{\rm mix}$.
The results are over plotted in figure~\ref{fig:euba} with the big open circles.
We modelled four cases of initial Eu abundance; $\eufe_{r} = 0.0, 0.5, 1.0$, and $1.5$ on the line of ``r-process only''.
We should note that the enrichemnt of Eu above is applicable only if $\eufe_{r} < \euba_{s} + \bah_{s} - \feoh = 0.8$ because the Eu abundance in the AGB envelope will not be depleted by the dredge-up of \sit-process elements.
The values of $F_{\rm mix}$ are set at $10^{-2}$, $10^{-1.3}$, $10^{-0.6}$, $10^{0.1}$, and $10^{0.8}$ from left to right for each sequence connected by the dashed lines.
We find that the ISM in Fornax must have pre-enriched by $\eufe > 0.5$ before the contribution of the \sit-process to the enrichment of the ISM starts to work at $\feoh \sim -1$.
Apparently, the ratio of ISM matter to AGB ejecta is in the range of $\sim 1 \hyp 20$.
This may be possible only if the production of \sit-process elments are efficient because the average value of $\eufe \sim 0.5$ is achieved at $\feoh \sim -1.5$ from the observations.

Other stars located in the middle of the \rit/\sit-dominant region in figure~\ref{fig:euba} are: five stars in Sculptor, four stars in Carina, one star in Draco, two stars in Sagittarius, one star in Leo I, and one star in Ursa Minor.
All these stars have similar Ba and Eu abundances with Galactic disk stars.

There are other two stars clearly located in the \sit-dominant region: SAGA\_UMi\_000005 (UMiK, \citet{Shetrone2001}) and SAGA\_Scl\_000007 (982, \citet{Geisler2005}).
It is not clear if they are carbon-enhanced stars because their carbon abundances have not been reported.

In the upper right part of figure~\ref{fig:euba}, the so-called CEMP-{\it r}/{\it s} (or CEMP-{\it i}, \cite{Hampel2016}) stars are clearly found among the Galactic stars (small points surrounded by the big circles in blue).
This was not the case in Paper II because seven out of nine stars located leftward of the solid line are newly added to the database.
Remaining two stars (CS 22948-027 and CS 29497-034) are the result of the conversion of abundance ratios using the adopted solar abundances.
In the particular case for these objects, the abundance conversions from \citet{Grevesse1996} or \citet{Grevesse1998} to \citet{Asplund2009} decreases barium abundance by 0.05 dex and europium abundance by 0.01 dex, which slightly shifts the star leftward on the diagram.
In any case, this class of stars are sensitive to the uncertainties associated with the abundance analyses.

\section{Summary}

We have developed the database for dwarf galaxies in the local group, which is the extension of the SAGA database for stars in the Milky Way (MW).
The database deals with the data of stellar abundances, photometry, stellar parameters, observational setups, equivalent width, adopted solar abundances, and bibliography.
We have compiled 24763 abundance data of 6109 unique stars from 25 galaxies, but most of them are based on low- and medium-resolution spectra.
We have also added metal-rich field stars in MW.
The database contains 10663 unique stars in total as of 2017 May.

The following updates have been applied to the new version of the database.
\begin{itemize}
\item Disk star sample have been included in the database by removing the selection criterion of papers that at least one of the stars should have $\feoh \leq -2.5$. This enables us to compare the stars in MW with those in dwarf galaxies.
\item Adopted solar abundances in the compiled papers are properly considered. This has partly resolved the inconsistency problem of the abundance data among the literature. Three datasets are provided for a different choice of scaled solar abundances. This can change the classification of stars, although its impact is minor.
\item The representative abundances for individual stars are defined. Users can choose average, median, or representative values from the data, where multiple data are available for elements and stellar parameters during the search. These data are used to classify the stars subject to abundances and evolutionary status.
\item We have added a new option for data plotting when the abundance data are reported by multiple papers. Users can plot all the data points or selected data from available papers for the same object, as well as a single data point selected according to the priority function as defined in the appendix~\ref{sec:priority}.
\end{itemize}

We have assigned star IDs in the database for stars in dwarf galaxies.
\begin{itemize}
\item The identification of stars are considered according to their position and brightness. We carefully checked the sensitivity on the criteria of identification and removed multiple counts for 2524 stars.
\item We renamed all the objects in the database, after applying the above identification process, with the unified format: {\it SAGA\_[Galaxy Name]\_[ID]}. This will reduce the confusion of star names in different galaxies.
\end{itemize}

We investigated the limitation of using a combined data set for elemental abundances taken from different papers.
We have to be more careful in comparing the abundances of stars in dwarf galaxies than in MW due to the usage of different analysis methods in deriving elemental abundances.
Comparisons are made for derived abundances for the same objects in different papers using different methods and/or different observational setups.
We find a discrepancy of abundances in several stars and a possible systematic difference among the literature values in some galaxies.
We compared the distance of individual stars using stellar parameters and brightness, and the distance to their host galaxies.
The disagreement of estimated distances can be, in most cases, ascribed to the uncertainties associated with derived stellar parameters, although some stars show significant discrepancies.

Using the extended database, we had several discussions on matters such as: metallicity distribution function, position distribution in each galaxy, radial metallicity gradient, fraction of carbon-enhanced stars, chemical enrichment of $\alpha$-elements, the comparison of abundances with the Galactic globular clusters, and the site of neutron-capture elements.

We introduced a cumulative metallicity distribution to discuss the star formation history of the galaxies.
Our simple interpretation of the slope of the cumulative metallicity distribution function is in reasonable agreement with other studies.
In spite of this agreement, we may need to consider the complex history of galaxy formation such as the inflow and outflow of gas.
A more homogeneous dataset will provide a more reliable discussion.

The radial metallicity gradients with direction dependence were examined using the position distribution.
Metallicity gradients are found in Fornax, Leo I, Draco, Sextans, and Leo II, and not in Carina as reported in the previous work.
We find no direction dependence in any of the galaxies registered in the database.

The fraction of carbon-enhanced metal-poor (CEMP) stars is apparently smaller in dwarf galaxies than in MW.
In particular, there are fewer stars with very large carbon enhancement, which is consistent with the paucity of CEMP stars with \sit-process element enhancement.
Only B\"ootes I stars show the similar trend of the CEMP fraction with MW around $\feoh \sim -2.5$, although the number of stars with measured carbon abundances is small.
Statistics are not sufficient for $\feoh < -3$, but the fraction of CEMP stars without the enhancement of \sit-process elements seems to be comparable to the Galactic stars as a whole.
A larger fraction of CEMP stars is found in ultra-faint dwarf galaxies than in classical dwarf spheroidal galaxies.
A scenario to explain the reason for this is discussed in \citet{Salvadori2015} where they ascribe the different CEMP fraction to the different metallicity distribution function.

We examined the difference in the shape of the metallicity distribution function, and determined the position of the so-called knee, both of which characterise the chemical evolution and star formation history of galaxies.
Our new technique to determine the position of the knees gives a consistent result with previous studies.

The abundance correlations between O and Na, and Mg and Al do not show similarities to those in the globular clusters in any of the galaxies, although the sample is too small to be robust.
Currently, we are not able to insist that any of the Galactic globular clusters are part of dwarf galaxies.

We confirmed that the abundances of neutron-capture elements in Fornax show a significant difference from the Galactic stars.
In particular, some Fornax stars comprise a different population in terms of Ba and Eu abundances, which indicate the pre-enrichment of the interstellar medium with the \rit-process by $\eufe > 0.5$ and the contamination with \sit-process yields at $\feoh \sim -1$.
This is the case if the efficiency of \sit-process element enrichment is high enough at this metallicity range in Fornax.

It will be important to collect more data on various elements with high resolution spectroscopy.
More data with new telescopes and projects on spectroscopic studies will improve our understanding of the formation of MW and the local group.

\begin{ack}
We thank the anonymous referee for his/her useful comments and suggestions, which improved the quality of the manuscript.
We are grateful to Yutaka Hirai for reading the manuscript and for giving helpful suggestions and comments on the chemical evolution and metallicity distribution function.
The manuscript has been improved by discussions with Joss Bland-Hawthorn, Charli Sakari, Matthew Shetrone, and Kim Venn during the conference, Galactic Archaeology and Stellar Physics 2016, held in Canberra.
This research made use of NASA's Astrophysics Data System, the Vizie-R astronomical database, operated at CDS Strasbourg, France. 
This research has been also made possible by all the efforts of those working on abundance analyses, together with the observations with low to high-dispersion spectroscopy by optical telescopes all over the world.
This work has been partially supported by a Grant-in-Aid for Scientific Research (JP23224004, JP268020, JP15HP7004, JP16K05287), from Japan Society of the Promotion of Science.
\end{ack}

\begin{appendix}
\section{Data selection priority}\label{sec:priority}
In plotting the data using the data retrieval subsystem, we sometimes have multiple data that satisfy search criteria.
If the search option is set to choose one of the data, the data selection is automatically optimised for the data of abundances, atmospheric parameters, photometric data, binary parameters, and isotopic ratios.
The data selection is determined by the priority parameter, $P$, which is calculated in the database according to the formula arbitrarily defined through our several tests to get reasonable results.
Because the definition of $P$ is not based on physics, it will be changed subject to future updates. 
The latest version of the function will be announced on the web site.

The following is the basic formula to determine the priority parameter, $P$:
\begin{equation}
P = (R / 10000)^{2} + ( Y - 2000 ).
\end{equation}
where $R \equiv \lambda / \Delta \lambda$ is the resolving power of the spectra and $Y$ the year of the publication of the paper.
The maximum value is adopted for the value of $R$ for the observing setups in the original paper.
If the $R$ is not given in the original papers, a typical value for observational setups is employed.

For abundance data, the priority parameter is determined by
\begin{equation}
P = (R / 10000)^{2} + ( Y - 2000 ) + S + \delta + E.
\end{equation}
where $S$ is a factor related to ionisation states or molecules, $\delta$ a factor related to the uncertainties of the value, and $E$ a factor related to the upper / lower limit to the value.
The value of $S$ is $0.1$ for neutral species and molecules, and is zero for species for which no information are available.
Temporarily, $S = 0.2$ for carbon abundance determined by CH lines to have a preference over C$_{2}$ lines.
The value of $\delta$ is set at $0.01 / \sigma$ where $\sigma$ is the total uncertainty or related value.
If $\sigma$ is not available in the literature, $\delta$ is set at $-0.05$.
The value of $E$ is set at $-10 \pm 0.1 | V |$ for the value $V$ of abundance with a lower [upper] limit.
If the value of $V$ does not include lower [upper] limit, $E = 0$.

\section{Identification of stars}\label{sec:id}

Our basic criteria for the identification of stars are (1) brightness difference within $0.4$ mag and (2) distance less than $3$ arcsecs.
However, the second condition is not sufficient to find the identical objects showing the minor difference in the coordinates in original papers.
To complete the identification of stars with large angular distance, we looked for other candidate stars around the objects in concern by relaxing the condition for separation to, say, $10$ arcsecs.
We found additional $537$ candidate pairs for identical objects.
Careful checks using SIMBAD made us to conclude that $49$ of them are identical objects.
The list of such stars are uploaded on the web site and are subject to be changed by future updates.

We should also consider the possibility that two different stars can be located within $3$ arcsecs.
We checked the star identification of two stars with a distance between $0.5$ and $3$ arcsecs using the SIMBAD database.
Our survey revealed $19$ such pairs in the database out of $493$ candidates, which are listed online.
It is to be noted that not all the stars are registered in SIMBAD.
It is difficult to ensure the identity of such stars in this case, even if the coordinates and brightness suggest stars in the SIMBAD database located within $0.5$ arcsecs distance with the difference in brightness less than $0.4$ mag in V and other bands.
Distant galaxies like Leo I tend to suffer from the misidentifications.
All the star IDs are available online and will be updated if necessary.

\end{appendix}

\bibliographystyle{apj}
\bibliography{apj-jour,reference,reference_mps_obs,reference_saga,reference_dsph,reference_solar}

\end{document}

%% file: Table1.tex
\begin{longtable}{l*{15}{r}}
  \caption{Number of registered stars in individual dwarf galaxies.
  }\label{tab:data}
\hline
\multicolumn{16}{c}{} \\
\endfirsthead
     Galaxy Name & N star & \feoh\ & \textrm{Ca} & \textrm{Ti} & \textrm{Si} & \textrm{Mg} & \textrm{C} & \textrm{Ba} & \textrm{Cr} & \textrm{Ni} & \textrm{Na} & \textrm{Nd} & \textrm{Eu} & \textrm{La} & $R \geq 15000$ \\
     \hline    
     Fornax        & 1700 & 1608 & 661 & 840 & 792 & 670 & 124 & 120 & 119 & 130 &  92 & 108 & 115 &  109 & 481 \\
     Carina        & 1229 & 1033 &  64 &  69 &  28 &  67 &   3 &  42 &  48 &  37 &  21 &  15 &  15 &  17 &  75 \\
     Leo I         &  904 &  904 & 796 & 749 & 644 & 531 &   0 &   2 &   1 &   2 &   2 &   2 &   2 &   2 &   2 \\
     Sculptor      &  533 &  533 & 437 & 387 & 373 & 310 & 300 &  24 &  18 &  21 &  25 &   7 &  19 &  10 & 106 \\
     Draco         &  346 &  340 & 200 & 154 & 178 &  91 & 203 &  16 &  14 &  14 &  14 &  10 &  12 &   0 &  16 \\
     Leo II        &  306 &  306 & 248 & 216 & 176 & 139 &   0 &   0 &   0 &   0 &   0 &   0 &   0 &   0 &   0 \\
     Sextans       &  301 &  301 &  96 &  76 &  73 &  43 &   3 &  13 &  13 &  11 &  12 &   6 &   3 &   1 &  13 \\
     Ursa Minor    &  226 &  225 & 131 & 113 & 128 &  78 &  93 &  21 &  21 &  11 &  18 &  17 &  17 &  11 &  18 \\
     CVn I         &  201 &  181 & 107 &  87 &  86 &  40 &   0 &   2 &   0 &   0 &   0 &   0 &   0 &   0 &   0 \\
     B\"ootes I      &   63 &   42 &  13 &  11 &   3 &  13 &  39 &  13 &  10 &  10 &  10 &   0 &   1 &   0 &  11 \\
     Sagittarius   &   61 &   56 &  45 &  44 &  14 &  45 &  11 &  14 &  11 &  14 &  13 &  14 &   2 &  14 &  61 \\
     Hercules      &   30 &   30 &  17 &   4 &   7 &   6 &   1 &  13 &   2 &   3 &   5 &   0 &   0 &   0 &  12 \\
     Ursa Major    &   23 &   17 &   9 &   8 &   7 &   2 &   0 &   0 &   0 &   0 &   0 &   0 &   0 &   0 &   0 \\
     Ursa Major II &   17 &    9 &   5 &   5 &   6 &   4 &   3 &   3 &   3 &   3 &   3 &   3 &   3 &   3 &   3 \\
     Triangulum II  &    15 &    6 &   6 &   4 &   4 &   5 &   2 &   2 &   1 &   1 &   2 &   1 &   1 &   1 &   3 \\
     Segue 1       &   11 &    8 &   7 &   7 &   7 &   7 &   8 &   7 &   7 &   7 &   7 &   0 &   7 &   1 &   7 \\
     Segue 2      &   10 &   10 &   6 &   6 &   8 &   3 &   0 &   0 &   0 &   0 &   0 &   0 &   0 &   0 &   0 \\
     Com Ber       &    9 &    9 &   7 &   4 &   5 &   3 &   3 &   3 &   3 &   3 &   3 &   3 &   3 &   3 &   3 \\
     Reticulum II  &    9 &    9 &   9 &   9 &   9 &   9 &   9 &   9 &   9 &   9 &   9 &   9 &   9 &   3 &   9 \\
     CVn II        &    8 &    8 &   4 &   5 &   2 &   2 &   0 &   1 &   0 &   0 &   1 &   0 &   0 &   0 &   0 \\
     B\"ootes II     &    6 &    6 &   4 &   4 &   4 &   4 &   4 &   4 &   4 &   4 &   4 &   4 &   0 &   0 &   4 \\
     Leo IV        &    5 &    5 &   4 &   5 &   4 &   3 &   1 &   3 &   1 &   1 &   3 &   0 &   0 &   0 &   1 \\
     Leo T         &    5 &    5 &   1 &   4 &   2 &   0 &   0 &   0 &   0 &   0 &   0 &   0 &   0 &   0 &   0 \\
     Tucana II     &    4 &    4 &   4 &   4 &   4 &   4 &   4 &   4 &   4 &   4 &   4 &   0 &   4 &   0 &   4 \\
     Willman I     &   14 &    8 &   0 &   0 &   0 &   0 &   0 &   0 &   0 &   0 &   0 &   0 &   0 &   0 &   0 \\
     \hline
\end{longtable}

%% file: Table2.tex
\begin{table}
  \tbl{Characterisctics of the knee position}{
    \begin{tabular}{*{5}{l}}
      \hline
      Galaxy & Knee position & Element & Slope & Other studies \\
      \hline
      Milky Way & $-1.0$ & Mg & $-0.367^{+0.0154}_{-0.0156} $ & \\
      Fornax    & $-2.1$ & Mg & $-0.22^{+0.059}_{-0.056} $ & $<-1.5$ (1), $-1.9$ (2) \\
      Sculptor  & $-2.1$ & Mg & $-0.44^{+0.16}_{-0.15} $ & $-1.8$ (1) \\
      Draco     & $-2.3$ & Ca & $-0.046^{+0.27}_{-0.22} $ & $-2.9$ (3) \\
      \hline
    \end{tabular}}\label{tab:slope}
    \begin{tabnote}
    (1) \citet{Tolstoy2009} \\
    (2) \citet{Hendricks2014a} \\
    (3) \citet{Cohen2009}
    \end{tabnote}
\end{table}